\documentclass[acmtog,nonacm,screen]{acmart}

\usepackage[utf8]{inputenc} 
\usepackage[T1]{fontenc}    
\usepackage{url}            
\usepackage{microtype}      

\usepackage{xcolor}        
\usepackage{pdfpages}
\usepackage{graphicx}
\usepackage{parskip}
\usepackage{booktabs}
\usepackage{listings}
\usepackage{float}
\usepackage{subcaption}
\usepackage{adjustbox}
\usepackage{tcolorbox}

\usepackage{listings} 
\usepackage{xcolor} 
\usepackage{xspace} 
\usepackage{graphicx} 
\usepackage{mdframed} 
\usepackage{lipsum} 

\usepackage{ifthen}
\usepackage{enumitem}
\usepackage{placeins}
\usepackage{afterpage}
\usepackage{subfiles}

\usepackage{amsthm}
\usepackage{amsmath}
\usepackage{nicefrac}        
\usepackage{siunitx}

\usepackage{tikz}
\usepackage{pgfplots}
\usepackage{pgfplotstable}
\pgfplotsset{compat=newest,scaled y ticks=true} 
\usepgfplotslibrary{units} 
\usetikzlibrary{spy}
\usetikzlibrary{backgrounds}
\usetikzlibrary{shapes,arrows,positioning,intersections,quotes}
\usetikzlibrary{matrix}
\usepackage{csvsimple}
\usepackage{graphbox} 

\usepackage{multirow}
\usepackage{tabularray}
\usepackage{tabulary}
\usepackage{colortbl}
\usepackage[referable]{threeparttablex}

\usepackage[switch,pagewise]{lineno} 

\newcommand{\cref}[2]{\hyperref[#2]{#1~\ref*{#2}}}
\newcommand{\colref}[3]{\hyperref[#2]{#1~\ref*{#2}{#3}}}
\newcommand{\figref}[1]{\cref{Fig.}{#1}}

\newcommand{\secref}[1]{\cref{Section}{#1}}

\newcommand{\tabref}[1]{\cref{Table}{#1}}

\newcommand{\figrefs}[2]{\hyperref[#1]{Figure~(\ref*{#1}--\ref*{#2})}}

\definecolor{codegreen}{rgb}{0,0.6,0}
\definecolor{codeblue}{rgb}{.11,.56,1}
\definecolor{codegray}{rgb}{0.5,0.5,0.5}
\definecolor{codepurple}{rgb}{0.58,0,0.82}
\definecolor{codeKeyword}{RGB}{211	54	130}
\definecolor{codeComment}{RGB}{42	161	152}
\definecolor{codeOmitted}{RGB}{108	113	196}
\definecolor{codeNumbers}{rgb}{0.5,0.5,0.5}
\definecolor{codeString}{RGB}{128, 161, 16}
\definecolor{textusercolor}{RGB}{0 0 0}
\definecolor{textgptcolor}{RGB}{0, 0, 0}
\definecolor{codebackcolour}{RGB}{	255  241 233}
\definecolor{backuserprompt}{RGB}{ 255, 247, 204}
\definecolor{backgptresponse}{RGB}{219 255 245}

\newcommand{\gpticon}{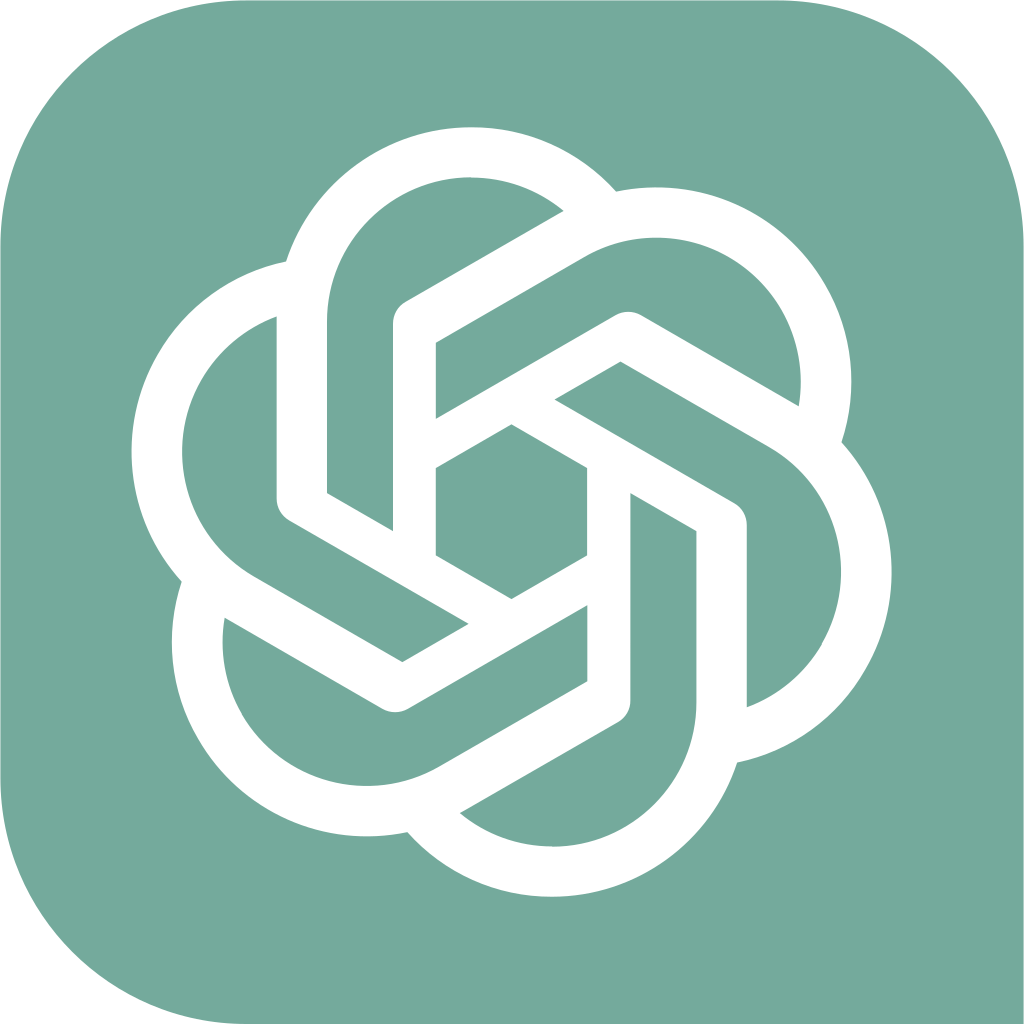}
\newcommand{\usericon}{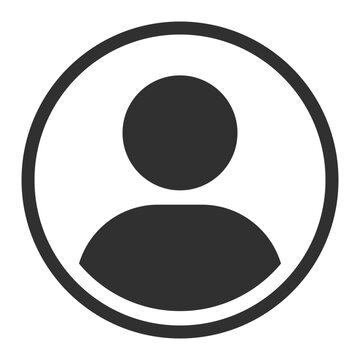}
\newcommand{\claudeicon}{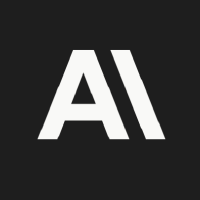}
\newcommand{\bardicon}{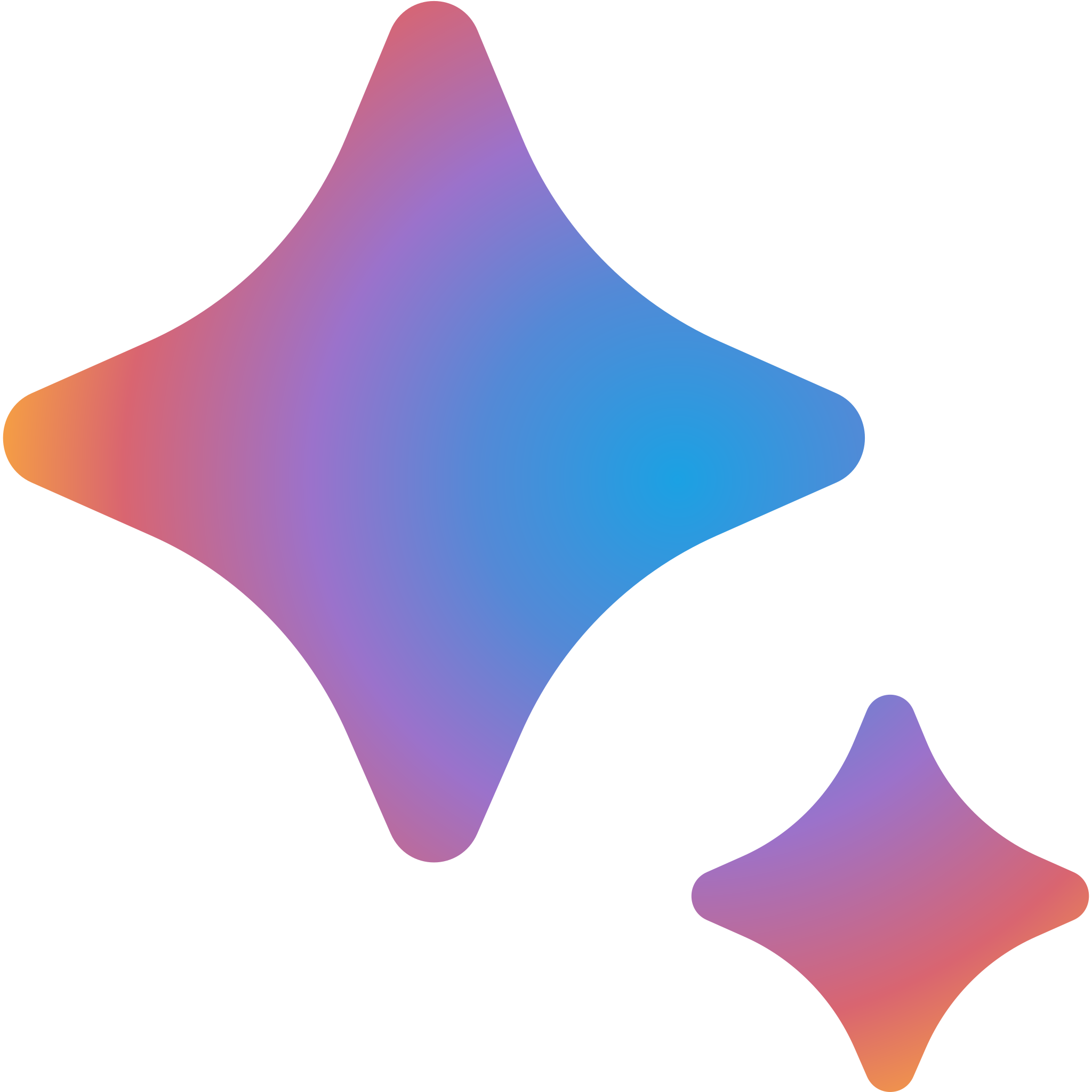}
\newcommand{\staricon}{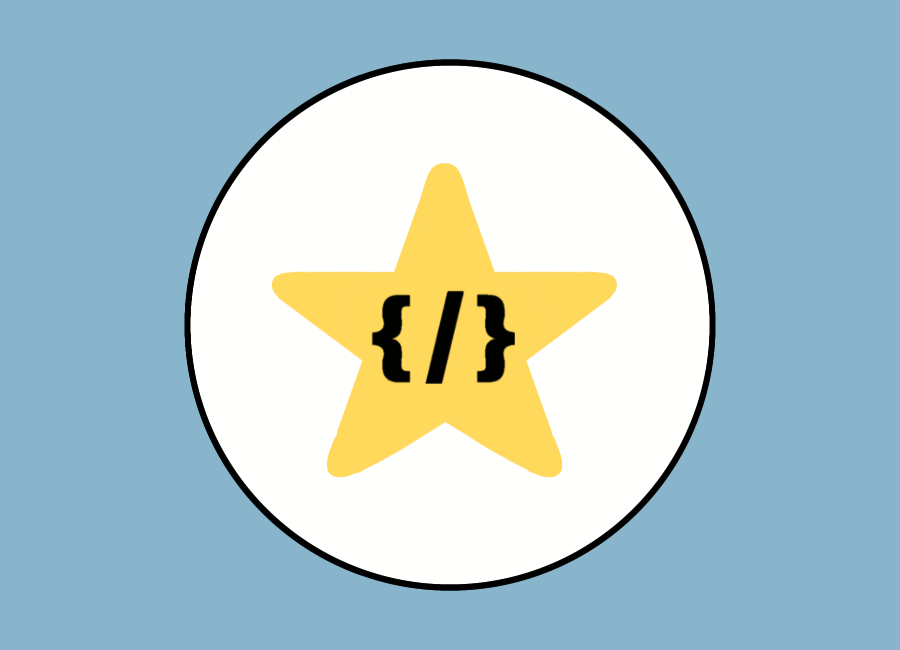}

\lstdefinelanguage{JavaScript}{
  keywords={typeof, new, true, false, catch, function, return, null, catch, switch, var, if, in, while, do, else, case, break, const},
  ndkeywords={class, export, boolean, throw, implements, import, this, require},
  sensitive=false,
  comment=[l]{//},
  morecomment=[s]{/*}{*/},
  morestring=[b]',
  morestring=[b]"
}

\lstdefinestyle{codestyle}{
    commentstyle=\color{codeComment},
    keywordstyle=\color{textusercolor},
    numberstyle=\tiny\color{codeNumbers},
    stringstyle=\color{codeString},
    basicstyle=\linespread{0.85}\footnotesize,
    columns=flexible,
    breakatwhitespace=false,         
    breaklines=true,                 
    captionpos=b,
    showspaces=false,
    showstringspaces=false,
    showtabs=false,
    tabsize=2,
    escapeinside={\$}{\$},
}

\surroundwithmdframed[
  hidealllines=true,
  backgroundcolor=codebackcolour,
  innerleftmargin=0pt,
  innertopmargin=0pt,
  innerbottommargin=0pt]{codeblock}

\newcommand\colboxcolor{codeComment} 
\newsavebox{\savedcolorbox}
\newenvironment{colbox}[2]
{
    \renewcommand\colboxcolor{#1}%
    \begin{lrbox}{\savedcolorbox}%
    \begin{minipage}{\dimexpr\linewidth-2\fboxsep\relax}
    \footnotesize
    \setlength{\parskip}{3pt}
    \bgroup\color{#2}
}
{
    \egroup\end{minipage}
    \end{lrbox}%
    \begin{center}
    \colorbox{\colboxcolor}{\usebox{\savedcolorbox}}
    \end{center}
}

\newsavebox{\savedfigurebox}
\newenvironment{blurbwithfig}[5]
{
    \newcommand{\figurewidth}{#1}
    \newcommand{\iconwidth}{0.06\linewidth}
    \newcommand{\blurbwidth}{0.982\linewidth - \figurewidth - \iconwidth}
    \newcommand{\imagetoshow}{#2}
    \newcommand{\backgroundcolor}{#3}
    \newcommand{\boxtextcolor}{#4}
    \newcommand{\icontoshow}{#5}

    \begin{lrbox}{\savedfigurebox}%
    \begin{minipage}[t]{\figurewidth}
        \vspace{2pt}
        \ifthenelse{\equal{\imagetoshow}{}}{}{\includegraphics[width=\linewidth]{\imagetoshow}}
    \end{minipage}\end{lrbox}%

    \noindent
    \begin{minipage}[t]{\iconwidth}
    \vspace{1pt}
    \centering
    \includegraphics[width=\linewidth]{\icontoshow}
    \end{minipage}
    \noindent
    \begin{minipage}[t]{\blurbwidth}
    \vspace{0pt}
    \begin{colbox}{\backgroundcolor}{\boxtextcolor}
}
{
    \end{colbox}
    \end{minipage}
    \hfill
    \usebox{\savedfigurebox}
}

\newenvironment{userprompt}[2]
{
    \begin{blurbwithfig}{#1}{#2}{backuserprompt}{textusercolor}{\usericon}
}
{
    \end{blurbwithfig}
}

\newenvironment{gptresponse}[2]
{
    \begin{blurbwithfig}{#1}{#2}{backgptresponse}{textgptcolor}{\gpticon}
}
{

    \end{blurbwithfig}
}

\newenvironment{clauderesponse}[2]
{
    \begin{blurbwithfig}{#1}{#2}{backgptresponse}{textgptcolor}{\claudeicon}
}
{

    \end{blurbwithfig}
}

\newenvironment{bardresponse}[2]
{
    \begin{blurbwithfig}{#1}{#2}{backgptresponse}{textgptcolor}{\bardicon}
}
{

    \end{blurbwithfig}
}

\newenvironment{starresponse}[2]
{
    \begin{blurbwithfig}{#1}{#2}{backgptresponse}{textgptcolor}{\staricon}
}
{

    \end{blurbwithfig}
}

\lstnewenvironment{codeblock}[1]
{   
    \lstset{style=codestyle} 
    \lstset{language=#1}
    \vspace{6pt}
}
{}

\newenvironment{chat}[1]
{
    \newcommand{\preventbreaks}{#1}
    \begin{center}
    \mdfsetup{nobreak=\preventbreaks}
    \begin{mdframed}[
        linecolor=black,
        innerleftmargin=2mm,
        innerrightmargin=0cm
        innertopmargin=0cm
        innerbottommargin=0cm
    ]{}

}
{ 
    \end{mdframed}
    \end{center}
}

\graphicspath{{Figures/}}
\DeclareGraphicsExtensions{.pdf,.png,.jpg}
\begin{document}
\bibliographystyle{unsrtnat}

\title{Towards Foundational AI Models for Additive Manufacturing: Language Models for G-Code Debugging, Manipulation, and Comprehension}

\author{Anushrut Jignasu}
\email{ajignasu@iastate.edu} 
\affiliation{%
  \institution{Iowa State University}
  \city{Ames}
  \state{Iowa}
  \country{USA}
}

\author{Kelly Marshall}
\email{km3888@nyu.edu} 
\affiliation{%
  \institution{New York University}
  \city{New York City}
  \state{New York}
  \country{USA}
}

\author{Baskar Ganapathysubramanian}
\email{baskarg@iastate.edu}
\affiliation{%
  \institution{Iowa State University}
  \city{Ames}
  \state{Iowa}
  \country{USA}
}

\author{Aditya Balu}
\email{baditya@iastate.edu} 
\affiliation{%
  \institution{Iowa State University}
  \city{Ames}
  \state{Iowa}
  \country{USA}
}

\author{Chinmay Hegde}
\email{chinmay.h@nyu.edu}
\authornotemark[1]
\affiliation{%
  \institution{New York University}
  \city{New York City}
  \state{New York}
  \country{USA}
}

\author{Adarsh Krishnamurthy}
\email{adarsh@iastate.edu} 
\authornote{Corresponding authors.}
\affiliation{%
  \institution{Iowa State University}
  \city{Ames}
  \state{Iowa}
  \country{USA}
}

\renewcommand{\shortauthors}{Jignasu et al.}

\begin{abstract}
3D printing or additive manufacturing is a revolutionary technology that enables the creation of physical objects from digital models. However, the quality and accuracy of 3D printing depend on the correctness and efficiency of the G-code, a low-level numerical control programming language that instructs 3D printers how to move and extrude material. Debugging G-code is a challenging task that requires a syntactic and semantic understanding of the G-code format and the geometry of the part to be printed. In this paper, we present the first extensive evaluation of six state-of-the-art foundational large language models (LLMs) for comprehending and debugging G-code files for 3D printing. We design effective prompts to enable pre-trained LLMs to understand and manipulate G-code and test their performance on various aspects of G-code debugging and manipulation, including detection and correction of common errors and the ability to perform geometric transformations. We analyze their strengths and weaknesses for understanding complete G-code files. We also discuss the implications and limitations of using LLMs for G-code comprehension.
\end{abstract}


\begin{teaserfigure}
    \centering
    \includegraphics[width=0.99\linewidth,clip,trim={0.0in 3.0in 0.0in 3.0in}]{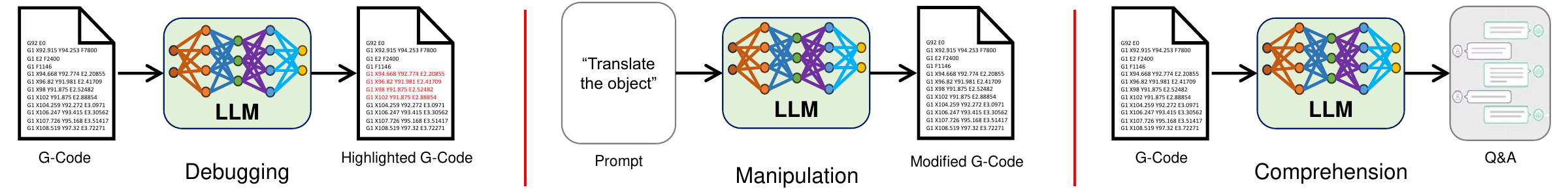}
    \caption{\sl Large language models (LLMs) have emerged as powerful tools in a variety of design applications. We present the first extensive evaluations of LLMs for G-code comprehension, debugging, and manipulation, analyze their strengths and weaknesses, and pave the way for future research.}
    \Description{This is the teaser figure for the article.}
    \label{fig:teaser}
\end{teaserfigure}

\maketitle

\section{Introduction}

The 3D printing revolution has democratized the easy, scalable, and efficient creation of physical objects from digital models. It has impacted applications in various domains, such as manufacturing, healthcare, construction, and the arts. However, the quality and accuracy of 3D printing depend on the correctness and efficiency of \emph{G-code}, a low-level numerical control programming language that instructs 3D printers how to move and extrude material. Dedicated software can generate the G-code for a particular part from the computer-aided design (CAD) model. Still, the efficacy of the generated G-code to correctly 3D print parts depends on extensive manual tuning of the G-code generation software. Consequently, generating high-quality G-code requires considerable expertise and is often an iterative process with manual debugging of the G-code itself.

Debugging or modifying the G-code to obtain a 3D printed part that is the same as the as-designed part is non-trivial. The process involves both syntactic and semantic aspects of the G-code, as well as geometric and physical constraints of the printing process. In addition, debugging the G-code is challenging once it is generated; since it is a low-level language, it is not very human-readable, and line-level commenting is typically absent (or rare at best). Consequently, G-code errors can be a common source of problems in 3D printing, as they can affect the quality and accuracy of the printed objects. Errors in G-code can result in defective prints, wasted resources, and damage to the 3D printer. These errors can damage the tool, the machine, or the print; in the worst case, code errors can even cause injury or human harm.

Generally, G-code errors can be classified into three types: syntax, configuration, or geometry errors. Syntax errors occur when the G-code commands do not follow the correct format or structure, such as missing parentheses, semicolons, or commas. These errors can prevent the G-code program from running or cause unexpected behavior. One of the common causes of syntax errors could be due to corruption of the G-code file itself. Since most G-code files are stored in ASCII format, simple manual opening and closing of the files could lead to certain lines being overwritten, deleted, or otherwise modified. Configuration errors occur when the machine settings do not match the G-code program requirements, such as using the wrong units, coordinate system, or tool offsets, which can result in inaccurate prints. Another kind of configuration error could occur if the 3D printer controller does not accept the G-code flavor. These configuration settings are usually stored on the G-code file header, allowing for easy inspection. However, this is still a tedious task since most of these configuration parameters are not in a human-readable form and require extensive manual reference checks. Geometry errors occur when the part orientation is wrong or if the coordinate systems of the part and the printer are not calibrated or aligned. These can also result from errors in the solid model (such as non-manifold surfaces) that can lead to errors in the slicing process. Such errors are more difficult to debug. Specifically, a geometry error can lead to collisions when the extrusion head comes into contact with an object it is not supposed to, such as hitting the bed, the print, or other parts of the printer, which can cause serious damage to the 3D printer itself.

While specific algorithmic solutions could be proposed for each of these categories of errors, they usually focus on specific aspects of the G-code debugging process. A flexible general solution to these challenges has emerged via the recent advances of foundational AI and Large language models (LLMs). These are powerful neural networks that can comprehend or generate natural language. They are trained on massive amounts of text data; moreover, they can be tuned to interpret, generate, and manipulate complex data types, such as code. Recent research has shown that natural language descriptions can be used for various tasks related to 3D printing, such as generating novel shapes~\citep{sanghi2022clip, marshall2023zeroforge, jain2022zero, lin2023magic3d}, editing scenes~\citep{haque2023instruct}, and reasoning about geometry in the volume space~\citep{kerr2023lerf}. Since G-code is a low-level language that directly instructs the 3D printing process at a fine-grained level, applying LLM technology for full G-code file comprehension is valuable and distinct from these related tasks. However, to the best of our knowledge, this has not been demonstrated in the literature\footnote{A possible exception is the very recent preprint~\citep{makatura2023can}, which performs a similar exploration of a specific LLM --- ChatGPT --- for building 3D CAD models, but they do not provide any significant results on evaluating LLMs for G-code.}, and we aim to bridge this gap.

Despite most of the G-code being algorithmically generated, G-code errors are still a prevalent concern. These errors can manifest in several ways, such as the geometry being misaligned from the bed, improper scaling, incorrect orientation, or uncalibrated process parameters not tailored to a specific printer. As a result, any attempt to debug or modify the G-code will require a basic understanding of the geometry being printed. To evaluate the applicability of using an LLM to perform these tasks, our evaluation metrics focus on debugging and applying geometric transformations directly to the G-code to counteract potential discrepancies. In addition, we also evaluate if the LLM can be used to geometrically reason about the part being printed directly from the G-code.

We present the first comprehensive evaluation of the state-of-the-art LLMs for debugging and modifying G-code files for extrusion-based additive manufacturing (AM). Given that G-code is a low-level language, LLMs struggle with handling the large number of lines or the size of the G-code file as input. Hence, we devise suitable approaches for LLMs to handle such low-level computer language data. We exclusively focus on pre-trained LLMs that do not require fine-tuning or domain adaptation; our main technical challenge is demonstrating their generalization abilities by framing (a suitable chain of) prompts for G-code manipulation. The main contributions of the paper are the following:
\begin{enumerate}
    \item We show how careful prompt engineering can enable pre-trained LLMs to understand G-codes for AM. We develop effective strategies for feeding large G-code files to context-length-limited LLMs.
    \item We systematically evaluate different state-of-the-art LLMs on a suite of comprehension tasks, including G-code debugging, capturing geometric transformations, and suggesting corrections. We compare the performance of different LLMs and analyze their strengths and weaknesses. 
    \item Best practices for using LLMs for dealing with such low-level (assembly-like) languages.
\end{enumerate}
We also discuss the implications and limitations of using LLMs for G-code debugging and comprehension, and finally, a few directions for future research.

\section{Background and Related Work}

\textbf{Additive Manufacturing (AM):} The emergence of robust AM techniques has paved the way for physically realizing Computer-Aided Design (CAD) models of varying complexity. Just as humans have evolved with language being the primary mode of information exchange, most AM techniques require a way to understand the model geometry, and this usually comes in the form of Geometric code (G-code). G-codes have been primarily developed for CNC machines but have been adapted for 3D printers. Printer-specific versions are more commonly seen in research communities since they allow for better local control, usually seen as printer-specific capabilities. Inherently, G-code does not contain any semantic information about the geometry being manufactured or about the geometry's intended function or material properties.

G-codes for additive manufacturing are collections of machine-specific commands that allow for building parts in an additive manner and serve as the intermediary between the digital representation (CAD) and AM hardware. They are usually generated by a slicing operation, often performed on tessellated versions of CAD geometries. These tessellated representations (stereolithography or STL file format) have been a universally adopted way of representing CAD models for 3D printing, owing to their simplicity~\citep{fadel1996accuracy}. The slicing operation generates layer information stored as coordinates that a 3D printer's nozzle follows. These coordinates are usually generated using plane-triangle intersection algorithms. Additionally, the slicing operation generates other necessary process parameters like extrusion values (for FDM 3D printing), infill patterns, and feed rates. The coordinates and process parameters are presented in the form of G-code. Each command typically begins with a letter like \textbf{G} or \textbf{M} and is usually followed by parameters. Each line of the G-code alters the state of the machine tool. The machine tool itself can be considered a state-space machine whose state is changed based on the instructions from the G-code. Using this abstraction, we can consider the G-code as a low-level machine language with characteristics closer to assembly languages. 

\textbf{Large Language Models:} Interest in the use of Large Language Models (LLMs), particularly in the research community, has increased exponentially in the last several months (as of summer 2023). The generative capabilities of these LLMs have yielded promising applications spanning a wide range of domains. Prominent LLMs include proprietary models like OpenAI's GPT series~\citep{brown2020language, openai2023gpt4} as well as LLMs which have been open-sourced for use by the general public such as Llama~\citep{touvron2023llama} and Starcoder~\citep{li2023starcoder}. While the training data for recent closed-source models have not been made public, we know that they are trained on large corpora of both natural language and code. However, these datasets encompass a vast array of domains, including science, arts, and engineering, providing the models with a comprehensive knowledge base. Such expansive training enables LLMs to generalize well across different tasks, even when only conditioned on textual prompts without further training.

Further work has extended LLM's applicability through the use of Reinforcement Learning from Human Feedback~\citep{ouyang2022training}, which fine-tunes language models on reward signals provided by human evaluators. This has spawned interactive versions of the aforementioned base models as well as some chat-specific competitors such as Google's Bard~\citep{Google} and Anthropic's Claude~\citep{Anthropic_2023}. These chatbots allow users to interface with an LLM's knowledge base and obtain useful answers that align with their preferences rather than simply receiving a prediction of the most likely text. 

Previous work by \citet{makatura2023can} offers an extensive study of GPT-4's capabilities for automating computational design and manufacturing pipelines. Their work shows that GPT-4 can reason about high-level properties and incorporate textually defined descriptions into the generation process but suffers from several limitations and a reliance on human feedback to correct its mistakes. 

Tokenization serves as the foundational layer in the architecture of LLMs, allowing the conversion of raw text into manageable units, called tokens, which have semantic meaning and are processed by the model. Tokenization methods have evolved significantly over time. Traditional methods represent decomposing input prompts into a sequence of character- or word-level tokens, Byte-pair Encoding (BPE)~\citep{sennrich2015neural}, WordPiece~\citep{schuster2012japanese}, and SentencePiece~\citep{kudo2018sentencepiece} have allowed for a transformative shift in tokenization. These methods effectively balance the granularity and expressivity of tokens, thus allowing for greater flexibility and reduced vocabulary size, ultimately enabling more efficient training and inference. Dealing with G-code is a challenging task from a tokenization perspective since G-code is a low-level language, and its interpretation is critical for accurate and efficient machine operation.

\textbf{Prompt engineering.} Despite their capabilities, LLMs face challenges due to the nuanced complexities of human language, where words can take on context-specific meanings. Effective prompting is, therefore, essential to guide LLMs' outputs in a more reliable and predictable manner. This process of ``prompt engineering'' involves designing task-specific language prompts to condition the LLM during inference. Techniques for prompt engineering range from automated methods~\citep{shin2020autoprompt, gao2020making} to manual approaches~\citep{reynolds2021prompt}. However, the scalability of gradient-based prompt-tuning techniques remains an open issue~\citep{lester2021power, liu2023gpt, qin2021learning}.

Several lines of inquiry in the literature address these challenges. For instance, research has explored the utility of fundamental contextual statements as reusable prompt patterns for various tasks~\citep{white2023prompt}. Other studies have proposed generating multiple prompts for a single task and aggregating the responses, employing weak supervision techniques for final predictions~\citep{arora2022ask}. A similar approach, known as decomposed prompting, breaks down complex tasks requiring chain-of-thought reasoning into simpler sub-tasks and uses individual LLMs specialized in each sub-task for more refined outputs~\citep{khot2022decomposed}. Additionally, the sensitivity of LLMs to the phrasing and structure of prompts has been analyzed to develop more effective single-prompt strategies~\citep{liu2021makes, zhao2021calibrate}. In our evaluations below, we leverage the technique of ``chain-of-thought''~\citep{wei2022chain} reasoning, a form of in-context learning used to precondition an LLM for G-code-related tasks.

\textbf{LLM technologies for manufacturing.} The use of LLMs in manufacturing is nascent and largely unexplored as of mid-2023. The closest work to our approach is a recent preprint by \citet{makatura2023can}, which makes a deep dive into how a specific LLM (ChatGPT) can be used in design and manufacturing applications. Another recent work by \citet{badini2023assessing} has looked into utilizing a specific LLM --- ChatGPT --- for AM process parameter optimization. They address printing-related issues, specifically for extrusion-based additive manufacturing using a soft material (Thermoplastic Polyurethane, TPU). They assess the utility of ChatGPT to modify the process parameters to reduce common errors such as warping, bed detachment, and stringing. However, it should be noted that they do not use the entire G-code file as input for modifying these parameters; these parameters are stored in the G-code header.

Our contributions are distinct and complementary to these existing works. We focus on assessing the capabilities of a diverse set of currently available LLMs for performing error correction, manipulating geometric shapes, and comprehending entire G-code files. Along the way, we address a key technical (and novel) challenge: dealing with limited context window lengths. We discuss details in the next section.

\section{Methods}
Our work aims to study the ability of LLMs to perform a range of operations. These include simple debugging, geometric transformations (such as translation, scaling, shearing, and rotation), and finally, geometry comprehension. In this section, we delve into our methodology to test the LLMs' ability in this area and evaluate their applicability for real-world use. 

Basic debugging evaluates the ability of the LLM to easily locate syntax errors. Geometric transformations change an object's position, orientation, and size in a predefined coordinate system. An LLM's ability to perform such transformations is a precursor to its ability to spatially understand, reason about, and manipulate an input geometry represented as G-code. Finally, we also use the LLM to reason about the geometry from the G-code alone. We utilize three canonical shapes for experimentation: a cube, a cylinder, and an S-shape. Each of the shapes was generated using Solidworks~\citep{SW2023}, sliced using PrusaSlicer~\citep{Prusa}, and visualized in Ultimaker's Cura~\citep{Cura}. 

We select the following LLM's for evaluation purposes\footnote{The list of high-capability LLMs is very rapidly growing as of mid-2023. We chose this particular subset of LLMs to reflect a diverse subset of the best available closed- and open-source models.}:
\begin{description}
    \item [GPT-3.5:] The most popular version of GPT developed by OpenAI; also the base version of ChatGPT.
    \item [GPT-4:] The current best version of GPT developed by OpenAI~\citep{openai2023gpt4}.
    \item [Bard:] An LLM using Google's PaLM 2 model~\citep{Google}.
    \item [Claude-2:] An LLM built by Anthropic~\citep{Anthropic_2023}.
    \item [Llama-2-70b:] Meta's open-source Llama-2 fine-tuned for chat (Llama-2-70b-chat-hf)~\citep{touvron2023llama} by HuggingFace and hosted on HuggingFace.
    \item [Starcoder:] The BigCode community's open-source Starcoder-Star\-chat-Beta~\citep{li2023starcoder}, fine-tuned for chat, and hosted on Huggingface.
\end{description}
All LLMs are evaluated with minimal hyperparameter tuning. Starcoder's interface on HuggingFace allows for manual tuning of \textit{temperature}, which influences the randomness of token selection; \textit{top-k tokens}, which limits the pool of tokens that can be sampled; \textit{top-p}, a parameter where higher values allow the model to sample more low-probability tokens; \textit{maximum new tokens}, which sets an upper limit on the number of tokens that can be generated in a single prompt; and a \textit{repetition penalty factor}, which discourages the model from repeating tokens or sequences. 

We adhere to default settings for all except \textit{maximum new tokens}, which we set as 1024. This adjustment allows us to generate more G-code sequences without being limited by token length. Furthermore, we make a concerted effort to use the same prompts for evaluation across all LLMs. However, owing to inherently different model architectures, attention mechanisms, potentially different training datasets, and varying token lengths of the LLMs we evaluate, such variance inherently affects the LLM's response to identical prompts, necessitating the occasional use of different prompts to achieve the same task across different models.

\begin{figure}[t!]
    \input{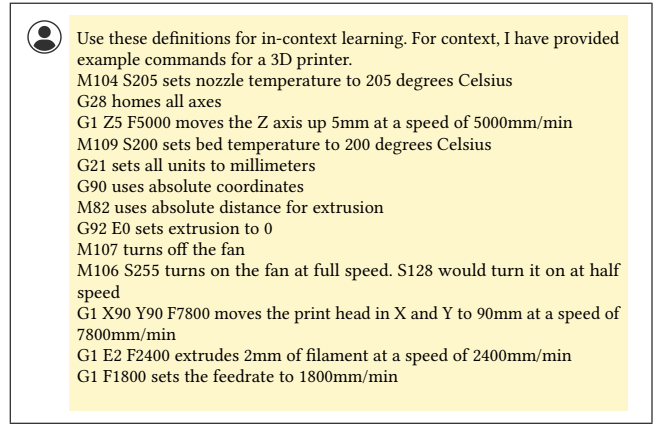}
    \caption{Initial prompt to each LLM.}
    \label{fig:G-code_def}
\end{figure}

In a general sense, all dialogues are performed with at least one of the following traits: 
\begin{itemize}[left=0pt]
    \item We begin each conversation with the same prompt (\figref{fig:G-code_def}).
    \item We provide chunks of G-code, owing to the varying token lengths. Empirically, we have found that a chunk length of 500 lines works for GPT-3.5 and GPT-4. For Bard, we use 350 lines. For Claude-2, Llama-2-70b, and Starcoder, we were able to provide the entire G-code as input.
    \item Depending on the evaluation metric, we provide G-code for the first layer on a conditional basis.
    \item We provide user feedback to solutions that are incomplete or omit key parts of the g-code. This is necessary to observe the support for iteration noted in~\citep{makatura2023can}. For these prompts, we maintain as much uniformity as possible between tasks and models. More details are provided in \secref{results_discuss}.
\end{itemize}

\textbf{Translation:}
This transformation involves moving every point of an object by a certain distance along a specified direction. It is important to note that this transformation does not affect the orientation or size of an object. It only shifts the location of the object. In our case, we translate in \textit{\textbf{X} by 10mm and in \textbf{Y} by 20mm} for all translation experiments. To translate the geometries represented by G-code, our primary prompt is ``\textit{Can you translate the shape by 10mm in X and 20mm in Y and return the updated G-code?}'' On a conditional basis, we also make use of the following secondary prompts - 
\begin{itemize}[left=0pt]
    \item ``\textit{Consider only the first layer of the G-code and translate it by 10mm in X and 20mm in Y and return the updated G-code}''
    \item ``\textit{Translate the G-code i sent at the beginning of our conversation by 10mm in X and 20mm in Y and return the updated G-code. include solid infill as well}''
\end{itemize}

\textbf{Scale:}
Scaling is a linear transformation, commonly used in 2D and 3D geometry processing. It enlarges or shrinks a geometry using a scale factor that is taken to be same along all three axes or unique to each axis. In this work, we select a \textit{uniform scaling factor of \textbf{2}} for all scaling experiments. We use this as our primary prompt - ``\textit{Can you scale the coordinates by a factor of 2 and give me the updated G-code?}'' and the following as our secondary prompts: 
\begin{itemize}[left=0pt]
    \item ``\textit{Can you scale the entire first layer by a factor of 2 and return the updated G-code?}''
\end{itemize}

\begin{table*}[t!]
    \centering
    \caption{Overview of capabilities we evaluate on for various LLMs. Color coding represents the average of performance across the shapes. \colorbox{green!25}{\phantom{X}}~depicts success, \colorbox{yellow!25}{\phantom{X}} depicts partial success, and \colorbox{red!25}{\phantom{X}} shows failure. \textit{Note: Debugging was only evaluated for the S-shape.}}
    \label{tab:overview}
    \setlength\extrarowheight{3pt}
    \newcolumntype{C}{>{\centering\arraybackslash}p{0.1\linewidth}}
    \newcolumntype{L}{>{\raggedright\arraybackslash}p{0.2\linewidth}}
    \begin{tabulary}{0.99\linewidth}{|L|C|C|C|C|C|C|}
        \hline
        \textbf{Capability} & \textbf{\texttt{GPT-3.5}} & \textbf{\texttt{GPT-4}} & \textbf{\texttt{Bard}} & \textbf{\texttt{Claude-2}} & \textbf{\texttt{Llama-2-70b}} & \textbf{Starcoder}\\
        \hline
        Debugging & \cellcolor{red!25} & \cellcolor{green!25} & \cellcolor{red!25} & \cellcolor{green!25} & \cellcolor{red!25} & \cellcolor{red!25}\\
        \hline
        Simple Transformations & \cellcolor{yellow!50} & \cellcolor{yellow!50} & \cellcolor{yellow!50} & \cellcolor{yellow!50} & \cellcolor{red!25} & \cellcolor{red!25}\\
        \hline
        Complex Transformations & \cellcolor{red!25}  & \cellcolor{yellow!50}  & \cellcolor{yellow!50}  & \cellcolor{red!25}  & \cellcolor{red!25} & \cellcolor{red!25}\\
        \hline
        Comprehension & \cellcolor{red!25} & \cellcolor{yellow!50} & \cellcolor{yellow!50} & \cellcolor{yellow!50} & \cellcolor{red!25} & \cellcolor{red!25}\\
        \hline
    \end{tabulary}
\end{table*}

\pagebreak

\textbf{Shear:}
Shearing is a common geometric transformation in 2D and 3D geometry processing. This operation amounts to skewing the original geometry without changing its area. For example, if we shear a 2D rectangle sideways, it becomes a parallelogram. The level of skewness is controlled using a \textit{shearing factor} to shift points parallel to the given direction by a distance perpendicular to their distance from the given line. 

In this work, we select a \textit{shearing factor} of \textbf{0.5} for all shearing experiments. All shearing experiments conducted in this work were provided with the following primary prompt - ``\textit{Can you shear the coordinates of the shape by a factor of 0.5 and return the updated G-code?}''. Depending on the output of the LLM, we also make use of the secondary prompts below on a conditional basis:
\begin{itemize}[left=0pt]
    \item ``\textit{Consider only the first layer of the G-code i sent. can you shear the coordinates by a factor of 0.5 and return the updated G-code?}''
    \item ``\textit{Consider only the first layer of the G-code, can you shear it by a factor of 0.5 along the positive x-direction and return the updated G-code?}''
\end{itemize}

\textbf{Rotation:}
Applying a rotation transformation involves turning every point on the object around a fixed point by a specified angle. Rotation is an axis-specific operation, and we select the \textbf{Z} axis (commonly used as the build direction for 3D printing) as our axis of rotation and \textit{rotate by 45 degrees} (counter-clockwise). We use the following prompt:
\begin{itemize}[left=0pt]
    \item ``\textit{I'd like you to produce a rotated version of the g-code shape I gave you and give me the g-code for this shape. I'd like to do a 45 degree counter-clockwise rotation along the z axis}''
\end{itemize}

\textbf{Comprehension:}
We define basic comprehension as the ability of the LLM to understand the underlying geometry and subsequently identify the specific 3D model being manufactured. Basic comprehension involves a nuanced understanding beyond simply parsing G-code commands or recognizing syntax. An LLM with strong comprehension abilities should be able to individually piece together commands and parameters to develop a coherent understanding of the manufactured model's shape, dimensions, and specific intricacies. As a first step, the LLM should be able to recognize common patterns in the G-code, such as boundaries, infills, holes, or curves. Comprehending such patterns is highly useful for debugging and manipulation tasks. We first begin by providing each LLM with G-code chunks of varying length (dependent on token length) and asking them to combine all chunks and analyze the shape being manufactured. Furthermore, depending on the success of the comprehension of the entire G-code, we also provide the G-code for the first layer and prompt again about the shape being manufactured. 

Given that chunk size varies across models, our comprehension prompts have minor differences. We use a mixture of the following prompts across the LLMs we test on:
\begin{itemize}
    \item ``\textit{combine all \textbf{T} chunks and tell me what shape does it represent?}''
    \item ``\textit{what shape is being manufactured by this gcode?}''
    \item ``\textit{can you analyze the first layer of the gcode to figure out the shape being manufactured and then figure out the dimensions of the shape?}''
    \item ``\textit{can you tell me what shape is represented by this gcode?}''
\end{itemize}

We provide detailed transcripts of all the chats with the LLMs in our GitHub repository.

\section{Results and Discussion}
\label{results_discuss}

In our evaluation of GPT-3.5, GPT-4, Bard, Claude-2, Llama-2-70b, and Starcoder, we have gained a significant understanding of their ability to reason about geometry, both semantically and syntactically. We have also witnessed their outputs' behavioral patterns when presented with a combination of varying prompts and the same G-codes. In this section, we provide an overview of their capabilities as seen via their ability to perform debugging~(\secref{debugging}), geometric transformations~(\secref{transformations}), and G-code comprehension~(\secref{comprehension}). Furthermore, \tabref{tab:overview} provides an overview of the capabilities we evaluate and the corresponding LLMs performance. We showcase the capabilities in increasing order of complexity. \textit{Simple Transformations} include translation and scaling, \textit{Complex Transformations} include shearing and rotation, and \textit{Comprehension} is about the LLM's ability to understand the underlying geometry. For a shape-specific performance visualization, we summarize the G-code results in \tabref{tab:transformations_overview}. We also include the longer transcripts of the chats in the Appendix \secref{Sec:Chats}.

\subsection{Simple Debugging}
\label{debugging}

We test out various LLM chatbots on a simple test case for line-by-line debugging. In this case, we take the 2D S-Shape (which is only 204 lines and therefore suitable for all context lengths) and replace the ``G1'' command in one line with ``G13'', an invalid G-code command.

\begin{figure}[t!]
    \input{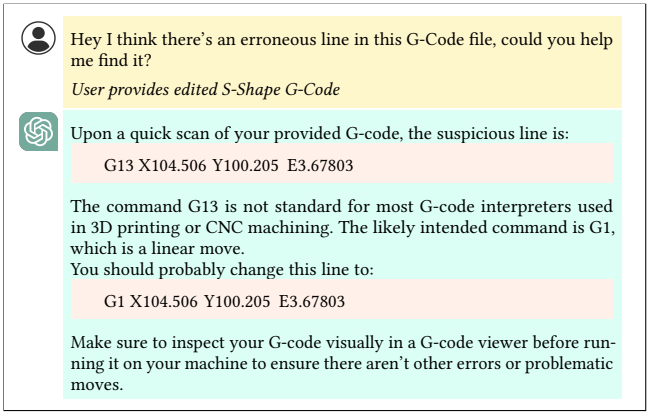}
    \caption{Prompting GPT-4 to find a bug in our G-Code}
    \label{fig:GPT4_easydebug}
\end{figure}

\begin{figure}[t!]
    \input{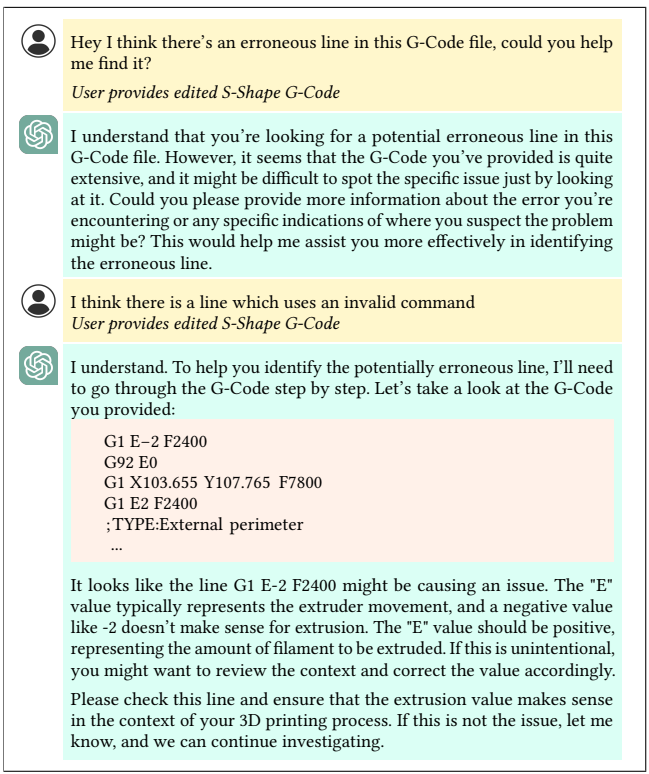}
    \caption{Prompting GPT-3.5 to find a bug in our G-Code}
    \label{fig:GPT3_easydebug}
\end{figure}

For each model, we first give it only the G-code file and ask it to look for an error. If the model understands the task but is unable to determine the cause of the error, we then provide an additional prompt saying that we are looking for an invalid command.

GPT-4 performs the best out of all the models, finding the error immediately and pointing it out, as shown in \figref{fig:GPT4_easydebug}. This is the ideal example of AI feedback, as it requires minimal user specification. \figref{fig:Claude_easydebug} displays Claude-2 also performing well, getting the correct answer after receiving the additional hint. In \figref{fig:GPT3_easydebug}, we see that GPT-3.5 is unable to solve the problem even with our additional prompting and suggests a candidate line that does not need fixing. While this is worse than GPT-4 and Claude-2's answers, GPT-3.5's chat responses show an awareness of its inability to provide a correct answer. In contrast, Bard immediately responds with a high degree of certainty about an incorrect presumed error, as seen in \figref{fig:Bard_easydebug}. This type of failure case is less desirable as it can mislead users who rely on AI support in debugging. StarCoder is entirely unable to comprehend the given task and does not seem to realize it is being given a G-Code debugging task, shown in \figref{fig:StarCoder_easydebug}. Llama-2-70b is unable to process the G-Code file as its context length is exceeded.

\begin{table*}[t!]
    \centering
    \caption{Overview of geometric transformation capabilities of the LLMs on different shapes. Color coding represents the performance. \colorbox{green!25}{\phantom{X}}~depicts success, \colorbox{yellow!25}{\phantom{X}}~depicts partial success, and \colorbox{red!25}{\phantom{X}}~shows failure.}
    \label{tab:transformations_overview}
    \setlength\extrarowheight{2pt}
    \newcolumntype{C}{>{\centering\arraybackslash}p{0.1\linewidth}}
    \newcolumntype{L}{>{\raggedright\arraybackslash}p{0.11\linewidth}}
    \begin{tabulary}{0.99\linewidth}{|L|L|C|C|C|C|C|C|}
        \hline
        \textbf{Shape} & \textbf{Capability} & \textbf{\texttt{GPT-3.5}} & \textbf{\texttt{GPT-4}} & \textbf{\texttt{Bard}} & \textbf{\texttt{Claude-2}} & \textbf{\texttt{Llama-2-70b}} & \textbf{Starcoder}\\
        \hline
        \multirow{4}{*}{Cube} & Translation & \cellcolor{green!25} & \cellcolor{green!25} & \cellcolor{yellow!50} & \cellcolor{green!25} & \cellcolor{red!25} & \cellcolor{red!25}\\
        \cline{2-7}
        & Scale &  \cellcolor{green!25} & \cellcolor{green!25} & \cellcolor{yellow!50} & \cellcolor{yellow!50} & \cellcolor{red!25} & \cellcolor{red!25}\\
        \cline{2-7}
        & Shear & \cellcolor{red!25} & \cellcolor{green!25} & \cellcolor{yellow!50} & \cellcolor{yellow!50} & \cellcolor{red!25} & \cellcolor{red!25}\\
        \cline{2-7}
        & Rotation & \cellcolor{yellow!50} & \cellcolor{red!25}  & \cellcolor{red!25}  & \cellcolor{red!25}  & \cellcolor{yellow!50 } & \cellcolor{red!25}\\
        \hline
        \multirow{4}{*}{Cylinder} & Translation & \cellcolor{yellow!50} & \cellcolor{yellow!50} & \cellcolor{yellow!50} & \cellcolor{green!25} & \cellcolor{red!25} & \cellcolor{red!25}\\
        \cline{2-7}
        & Scale & \cellcolor{red!25}  & \cellcolor{yellow!50}  & \cellcolor{yellow!50}  & \cellcolor{yellow!50}  & \cellcolor{red!25} & \cellcolor{red!25}\\
        \cline{2-7}
        & Shear & \cellcolor{red!25} & \cellcolor{green!25} & \cellcolor{yellow!50} & \cellcolor{yellow!50} & \cellcolor{red!25} & \cellcolor{red!25}\\
        \cline{2-7}
        & Rotation & \cellcolor{red!25} & \cellcolor{yellow!50} & \cellcolor{red!25} & \cellcolor{red!25} & \cellcolor{red!25} & \cellcolor{red!25}\\
        \hline
        \multirow{4}{*}{S-shape} &Translation & \cellcolor{yellow!50} & \cellcolor{yellow!50} & \cellcolor{yellow!50} & \cellcolor{yellow!50} & \cellcolor{red!25} & \cellcolor{red!25}\\
        \cline{2-7}
        & Scale & \cellcolor{yellow!50}  & \cellcolor{yellow!50}  & \cellcolor{red!25}  & \cellcolor{red!25} & \cellcolor{red!25} & \cellcolor{red!25}\\
        \cline{2-7}
        & Shear & \cellcolor{yellow!50} & \cellcolor{yellow!50} & \cellcolor{yellow!50} & \cellcolor{red!25} & \cellcolor{red!25} & \cellcolor{red!25}\\
        \cline{2-7}
        & Rotation & \cellcolor{red!25} & \cellcolor{red!25} & \cellcolor{red!25} & \cellcolor{red!25} & \cellcolor{red!25} & \cellcolor{red!25}\\
        \hline
    \end{tabulary}
\end{table*}

\subsection{Geometric Transformations}
\label{transformations}
To assess the ability of our LLM test suite on common transformation operations, we conduct dialogues that ask them to perform translation, scaling, shearing, and rotation on the provided G-code. Empirically, we find that due to the varying token lengths across all the LLMs we test, the geometric transformations may or may not be applied to the entire G-code. A possible reason for this may be that the token length limits may be the bottleneck for an LLM's spatial reasoning capability. This also limits the LLM's ability to infer the shape being manufactured. For this reason, we also test their performance on G-code for a single layer. We found that prompts asking for evaluation on a single G-code layer received far more engaging and positive results than prompts asking for evaluation on the entire G-code. It is important to note that due to the order of transformation related prompts, in some cases, the LLMs refuse to acknowledge the fact that the transformation from the current prompt is to be applied on the original G-code instead of the G-code generated by the preceding prompt. In such cases, we explicitly provide G-code for the first layer and repeat the prompt. 

In the G-code visualization figures, some subfigures that do not fit are shown in the Appendix. Figures with a cross mark are cases where the LLM failed to generate G-code, generated G-code with incorrect structure or missing \textit{E} (extrusion) values, or failed to understand the represented shape. More information about such cases is provided in respective figure captions. It is difficult to visually convey the change in G-code after geometric transformations. For that reason, we use a grid background, which aids in visualizing the relative change from the original G-code.

\subsubsection{Translation}\label{Sec:ResultsTranslation}
\begin{figure}[t!]
    \centering
    \small
    \begin{tblr}
    {
      colspec = {X[{0.2\linewidth},c,t]X[{0.19\linewidth},c,t]X[{0.19\linewidth},c,t]X[{0.19\linewidth},c,t]},
      stretch = 0,
      rowsep = 8pt,
      hlines = {black, 1pt},
      vlines = {black, 1pt},
    }
    \texttt{Expected} &
    \includegraphics[align=c,width=0.95\linewidth,clip,trim={0.0in 0.0in 0.0in 0.0in}]{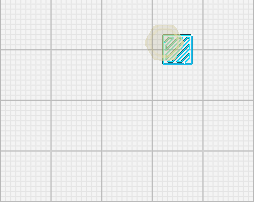} &
    \includegraphics[align=c,width=0.95\linewidth,clip,trim={0.0in 0.0in 0.0in 0.0in}]{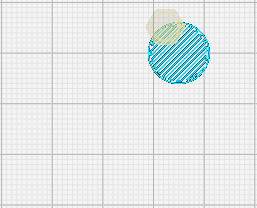} &
    \includegraphics[align=c,width=0.95\linewidth,clip,trim={0.0in 0.0in 0.0in 0.0in}]{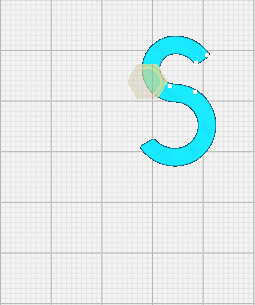}\\
    
    \texttt{GPT-3.5} &
    \includegraphics[align=c,width=0.95\linewidth,clip,trim={0.0in 0.0in 0.0in 0.0in}]{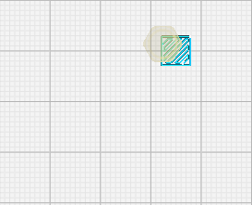} &
    \includegraphics[align=c,width=0.95\linewidth,clip,trim={0.0in 0.0in 0.0in 0.0in}]{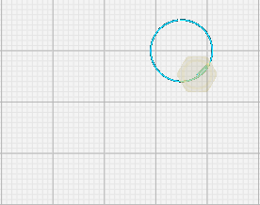} &
    \includegraphics[align=c,width=0.95\linewidth,clip,trim={0.0in 0.0in 0.0in 0.0in}]{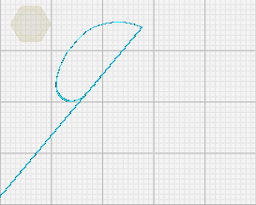}\\

    \texttt{GPT-4} & 
    \includegraphics[align=c,width=0.95\linewidth,clip,trim={0.0in 0.0in 0.0in 0.0in}]{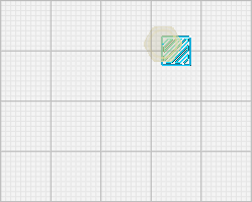} &
    \includegraphics[align=c,width=0.95\linewidth,clip,trim={0.0in 0.0in 0.0in 0.0in}]{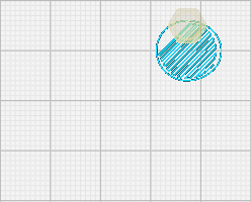} &
    \includegraphics[align=c,width=0.95\linewidth,clip,trim={0.0in 0.0in 0.0in 0.0in}]{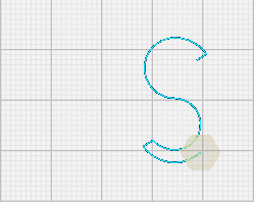}\\

    \texttt{Bard} &
    \includegraphics[align=c,width=0.95\linewidth,clip,trim={0.0in 0.0in 0.0in 0.0in}]{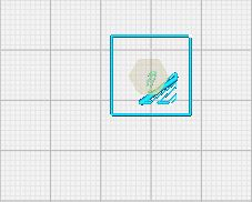} &
    \includegraphics[align=c,width=0.95\linewidth,clip,trim={0.0in 0.0in 0.0in 0.0in}]{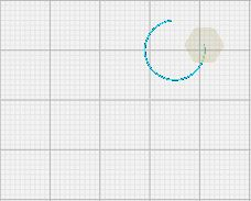} &
    \includegraphics[align=c,width=0.95\linewidth,clip,trim={0.0in 0.0in 0.0in 0.0in}]{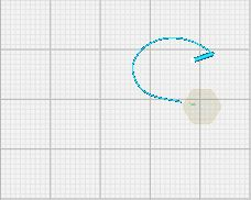}\\

    \texttt{Claude-2} &
    \includegraphics[align=c,width=0.95\linewidth,clip,trim={0.0in 0.0in 0.0in 0.0in}]{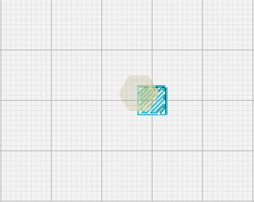} &
    \includegraphics[align=c,width=0.95\linewidth,clip,trim={0.0in 0.0in 0.0in 0.0in}]{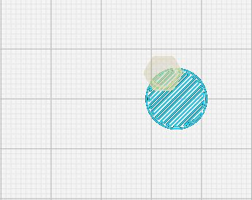} &
    \includegraphics[align=c,width=0.95\linewidth,clip,trim={0.0in 0.0in 0.0in 0.0in}]{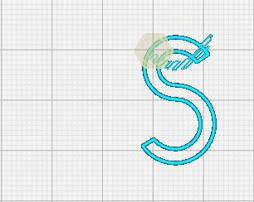}\\

    \texttt{Llama-2-70b} &
    \includegraphics[align=c,width=0.95\linewidth,clip,trim={0.0in 0.0in 0.0in 0.0in}]{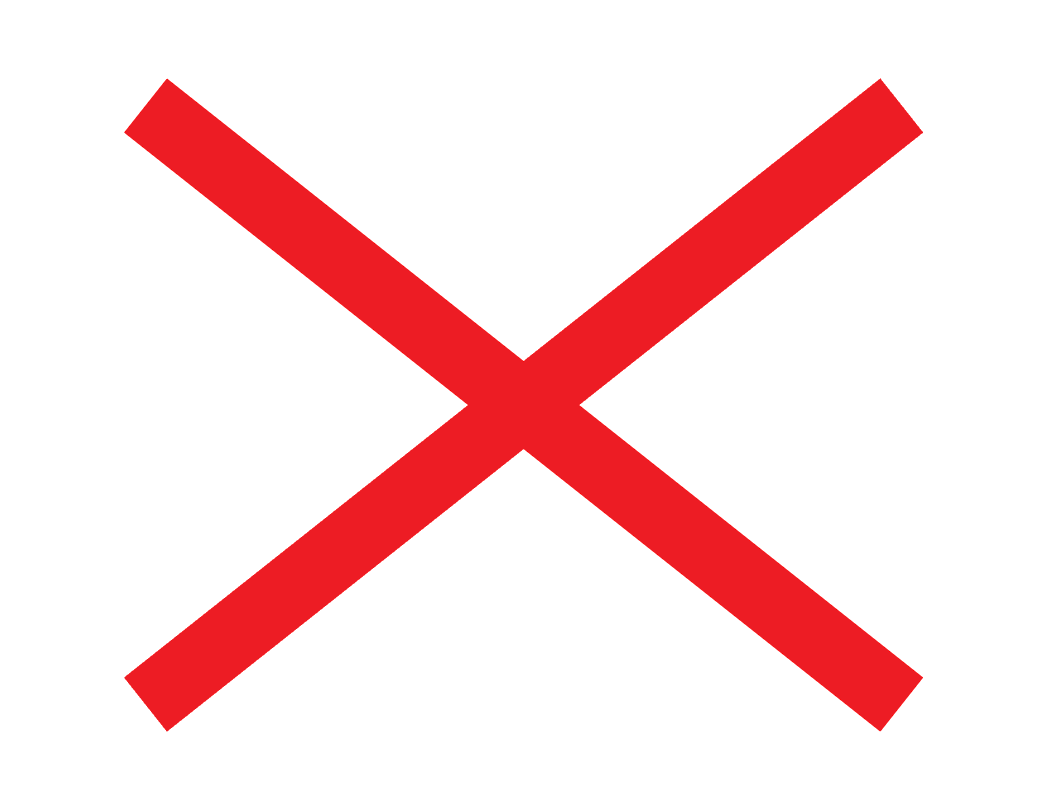} &
    \includegraphics[align=c,width=0.95\linewidth,clip,trim={0.0in 0.0in 0.0in 0.0in}]{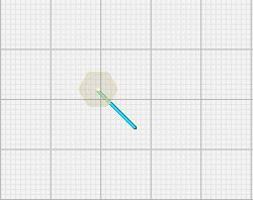} &
    \includegraphics[align=c,width=0.95\linewidth,clip,trim={0.0in 0.0in 0.0in 0.0in}]{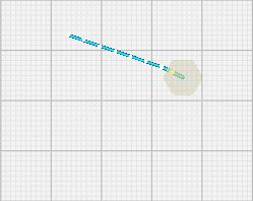}\\
    
    \texttt{Starcoder} &
    \includegraphics[align=c,width=0.95\linewidth,clip,trim={0.0in 0.0in 0.0in 0.0in}]{Figures/png/cross_red.png} &
    \includegraphics[align=c,width=0.95\linewidth,clip,trim={0.0in 0.0in 0.0in 0.0in}]{Figures/png/cross_red.png} &
    \includegraphics[align=c,width=0.95\linewidth,clip,trim={0.0in 0.0in 0.0in 0.0in}]{Figures/png/cross_red.png}\\
\end{tblr}
\caption{G-code visualization for translation operation on all LLMs. Expected G-code (top row). GPT-3.5 (second row), GPT-4 (third row), Bard (fourth row), Claude-2 (fifth row), Llama-2-70b (sixth row), Starcoder (last row).}
\label{fig:translation_G-code_vis}
\end{figure}

A conversation with Claude-2 is shown in \figref{fig:Claude2_translate_cube}. The output G-code visualizations for all translation operations are shown in \figref{fig:translation_G-code_vis}. From the G-code visualizations, we observe that GPT-3.5 generates a diagonal extrusion line that goes to the origin of the print bed. GPT-4 cannot generate the entire infill for the cylinder and generate the boundary and infill for the S-shape. Bard also struggles with shape boundary completion and infill generation. Claude-2 can generate entire boundaries of the shape but falls short during infill generation for the S-shape. Llama-2-70b fails to generate a correct G-code structure for the cube and does not understand the cylinder and S-shape. Starcoder fails to understand the shape represented by G-code for cube and cylinder. For the S-shape, it generates a G-code with an incorrect structure.

\pagebreak

\subsubsection{Scale}\label{Sec:ResultsScale}
\begin{figure}[t!]
    \centering
    \small
    \begin{tblr}
    {
      colspec = {X[{0.2\linewidth},c,t]X[{0.19\linewidth},c,t]X[{0.19\linewidth},c,t]X[{0.19\linewidth},c,t]},
      stretch = 0,
      rowsep = 6pt,
      hlines = {black, 1pt},
      vlines = {black, 1pt},
    }
    \texttt{Expected} &
    \includegraphics[align=c,width=0.95\linewidth,clip,trim={0.0in 0.0in 0.0in 0.0in}]{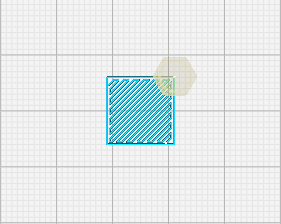} &
    \includegraphics[align=c,width=0.95\linewidth,clip,trim={0.0in 0.0in 0.0in 0.0in}]{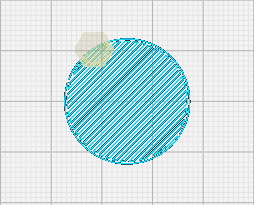} &
    \includegraphics[align=c,width=0.95\linewidth,clip,trim={0.0in 0.0in 0.0in 0.0in}]{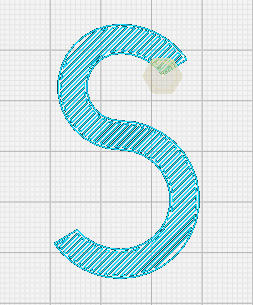}\\
    
    \texttt{GPT-3.5} &
    \includegraphics[align=c,width=0.95\linewidth,clip,trim={0.0in 0.0in 0.0in 0.0in}]{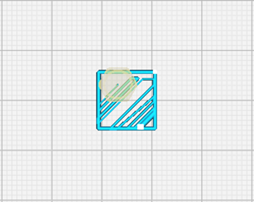} &
    See \figref{fig:outlier_scale_cylinder_gpt35} &
    \includegraphics[align=c,width=0.95\linewidth,clip,trim={0.0in 0.0in 0.0in 0.0in}]{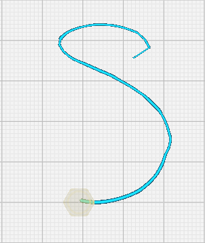}\\

    \texttt{GPT-4} & 
    \includegraphics[align=c,width=0.95\linewidth,clip,trim={0.0in 0.0in 0.0in 0.0in}]{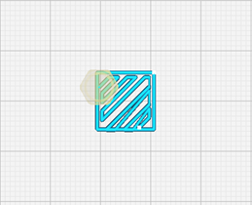} &
    \includegraphics[align=c,width=0.95\linewidth,clip,trim={0.0in 0.0in 0.0in 0.0in}]{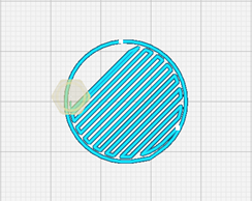} &
    \includegraphics[align=c,width=0.95\linewidth,clip,trim={0.0in 0.0in 0.0in 0.0in}]{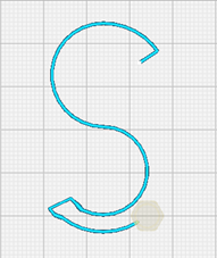}\\

    \texttt{Bard} &
    \includegraphics[align=c,width=0.95\linewidth,clip,trim={0.0in 0.0in 0.0in 0.0in}]{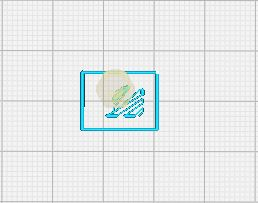} &
    \includegraphics[align=c,width=0.95\linewidth,clip,trim={0.0in 0.0in 0.0in 0.0in}]{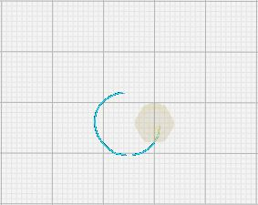} &
    See \figref{fig:outlier_scale_s_bard} \\

    \texttt{Claude-2} &
    \includegraphics[align=c,width=0.95\linewidth,clip,trim={0.0in 0.0in 0.0in 0.0in}]{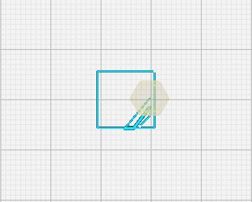} &
    \includegraphics[align=c,width=0.95\linewidth,clip,trim={0.0in 0.0in 0.0in 0.0in}]{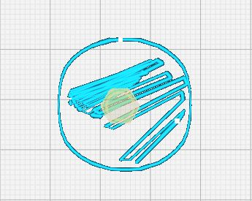} &
    See \figref{fig:outlier_scale_s_claude2}\\

    \texttt{Llama-2-70b} &
    \includegraphics[align=c,width=0.95\linewidth,clip,trim={0.0in 0.0in 0.0in 0.0in}]{Figures/png/cross_red.png} &
    See \figref{fig:outlier_scale_cylinder_llama2} &
    \includegraphics[align=c,width=0.95\linewidth,clip,trim={0.0in 0.0in 0.0in 0.0in}]{Figures/png/cross_red.png}\\
    
    \texttt{Starcoder} &
    \includegraphics[align=c,width=0.95\linewidth,clip,trim={0.0in 0.0in 0.0in 0.0in}]{Figures/png/cross_red.png} &
    \includegraphics[align=c,width=0.95\linewidth,clip,trim={0.0in 0.0in 0.0in 0.0in}]{Figures/png/cross_red.png} &
    \includegraphics[align=c,width=0.95\linewidth,clip,trim={0.0in 0.0in 0.0in 0.0in}]{Figures/png/cross_red.png}\\
\end{tblr}
\caption{G-code visualization for scaling operation on all LLMs. Expected G-code (top row). GPT-3.5 (second row), GPT-4 (third row), Bard (fourth row), Claude-2 (fifth row), Llama-2-70b (sixth row), Starcoder (last row).}
\label{fig:scale_G-code_vis}
\end{figure}

An example conversation with GPT-4 is shown in \figref{fig:GPT4_scale_cylinder}. The output G-code visualizations for scaling are shown in \figref{fig:scale_G-code_vis}. The G-code visualizations show that GPT-3.5 cannot capture the entire S-shape. GPT-4 can capture most of the first layer for the cylinder and cube but generates an incomplete S-shape. Bard uses the output of the previous operation (shearing) and scales that to generate a rectangular-looking boundary. For the cylinder, Bard generates an incomplete boundary. Claude-2 is not able to generate infill for the sheared cube and cylinder. Llama-2-70b generates a G-code with an incorrect structure for the cube and S-shape. Starcoder fails to understand the shape of the cube and cylinder and generates an incorrect G-code for the S-shape.

\subsubsection{Shear}\label{Sec:ResultsShear}
\begin{figure}[t!]
    \centering
    \small
    \begin{tblr}
    {
      colspec = {X[{0.2\linewidth},c,t]X[{0.19\linewidth},c,t]X[{0.19\linewidth},c,t]X[{0.19\linewidth},c,t]},
      stretch = 0,
      rowsep = 6pt,
      hlines = {black, 1pt},
      vlines = {black, 1pt},
    }
    \texttt{Expected} &
    \includegraphics[align=c,width=0.95\linewidth,clip,trim={0.0in 0.0in 0.0in 0.0in}]{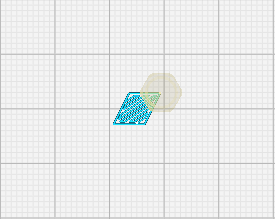} &
    \includegraphics[align=c,width=0.95\linewidth,clip,trim={0.0in 0.0in 0.0in 0.0in}]{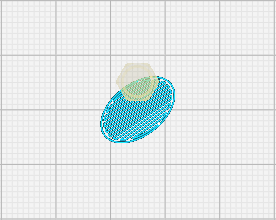} &
    \includegraphics[align=c,width=0.95\linewidth,clip,trim={0.0in 0.0in 0.0in 0.0in}]{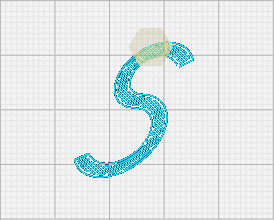}\\
    
    \texttt{GPT-3.5} &
    \includegraphics[align=c,width=0.95\linewidth,clip,trim={0.0in 0.0in 0.0in 0.0in}]{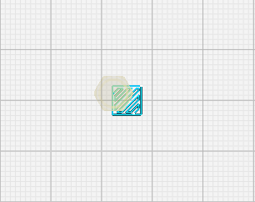} &
    \includegraphics[align=c,width=0.95\linewidth,clip,trim={0.0in 0.0in 0.0in 0.0in}]{Figures/png/cross_red.png} &
    \includegraphics[align=c,width=0.95\linewidth,clip,trim={0.0in 0.0in 0.0in 0.0in}]{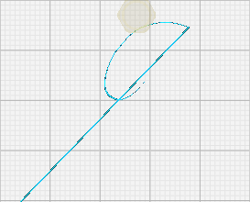}\\

    \texttt{GPT-4} & 
    \includegraphics[align=c,width=0.95\linewidth,clip,trim={0.0in 0.0in 0.0in 0.0in}]{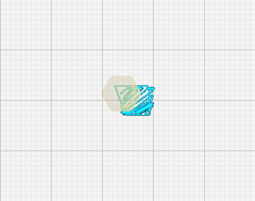} &
    \includegraphics[align=c,width=0.95\linewidth,clip,trim={0.0in 0.0in 0.0in 0.0in}]{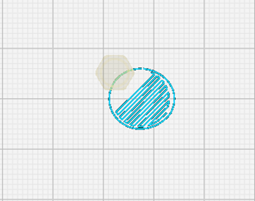} &
    \includegraphics[align=c,width=0.95\linewidth,clip,trim={0.0in 0.0in 0.0in 0.0in}]{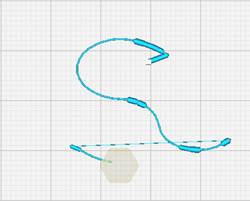}\\

    \texttt{Bard} &
    \includegraphics[align=c,width=0.95\linewidth,clip,trim={0.0in 0.0in 0.0in 0.0in}]{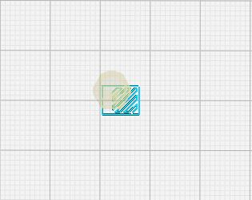} &
    \includegraphics[align=c,width=0.95\linewidth,clip,trim={0.0in 0.0in 0.0in 0.0in}]{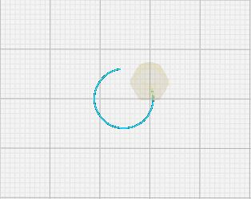} &
    \includegraphics[align=c,width=0.95\linewidth,clip,trim={0.0in 0.0in 0.0in 0.0in}]{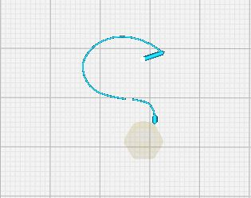}\\

    \texttt{Claude-2} &
    \includegraphics[align=c,width=0.95\linewidth,clip,trim={0.0in 0.0in 0.0in 0.0in}]{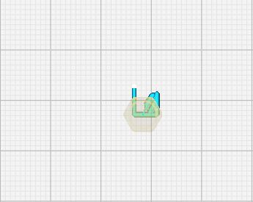} &
    \includegraphics[align=c,width=0.95\linewidth,clip,trim={0.0in 0.0in 0.0in 0.0in}]{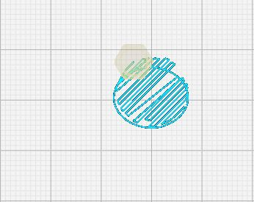} &
    See \figref{fig:outlier_shear_s_claude2} \\

    \texttt{Llama-2-70b} &
    \includegraphics[align=c,width=0.95\linewidth,clip,trim={0.0in 0.0in 0.0in 0.0in}]{Figures/png/cross_red.png} &
    \includegraphics[align=c,width=0.95\linewidth,clip,trim={0.0in 0.0in 0.0in 0.0in}]{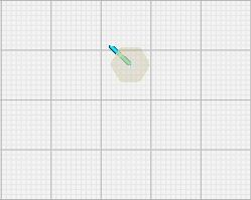} &
    \includegraphics[align=c,width=0.95\linewidth,clip,trim={0.0in 0.0in 0.0in 0.0in}]{Figures/png/cross_red.png}\\
    
    \texttt{Starcoder} &
    \includegraphics[align=c,width=0.95\linewidth,clip,trim={0.0in 0.0in 0.0in 0.0in}]{Figures/png/cross_red.png} &
    \includegraphics[align=c,width=0.95\linewidth,clip,trim={0.0in 0.0in 0.0in 0.0in}]{Figures/png/cross_red.png} &
    \includegraphics[align=c,width=0.95\linewidth,clip,trim={0.0in 0.0in 0.0in 0.0in}]{Figures/png/cross_red.png}\\
\end{tblr}
\caption{G-code visualization for shearing operation on all LLMs. Expected G-code (top row). GPT-3.5 (second row), GPT-4 (third row), Bard (fourth row), Claude-2 (fifth row), Llama-2-70b (sixth row), Starcoder (last row).}
\label{fig:shear_G-code_vis}
\end{figure}

Shearing is a relatively complex transformation as the LLMs have to not only scale the coordinates along a direction but also add them. An example conversation with Bard is shown in \figref{fig:Bard_shear_s}. The output G-code visualizations for shearing are shown in \figref{fig:shear_G-code_vis}. The results show that GPT-3.5 generates a similar-looking cube after shearing and fails to generate a sheared version of the cylinder. Additionally, it struggles with shape completion for a sheared S-shape. GPT-4 can recover most of the boundary for the cube and cylinder, but it finds it difficult to generate infill and complete the S-shape. Bard elongates the cube but is not able to complete the infill. It also struggles with completing the boundary for the cylinder and S-shape. Claude-2 significantly struggles with shearing the cube and S-shape. However, performs marginally better than the cylinder by completing the boundary as well as infill. Llama-2-70b generates an incorrect G-code structure for the cube and S-shape and loses context while shearing the cylinder. Starcoder fails again to understand and generate any meaningful G-code for visualization.

\subsubsection{Rotation}
Rotation is more challenging than the previously listed operations. While it can still be implemented using line-wise transformations of the original G-code, these transformations require multiplying the specified coordinates by a rotation matrix calculated based on the rotation angle. An example conversation is shown in \figref{fig:Bard_rotation_chat}

\begin{figure}[t!]
    \centering
    \small
    \begin{tblr}
    {
      colspec = {X[{0.2\linewidth},c,t]X[{0.19\linewidth},c,t]X[{0.19\linewidth},c,t]X[{0.19\linewidth},c,t]},
      stretch = 0,
      rowsep = 6pt,
      hlines = {black, 1pt},
      vlines = {black, 1pt},
    }
    \texttt{Expected} &
    \includegraphics[align=c,width=0.95\linewidth,clip,trim={0.0in 0.0in 0.0in 0.0in}]{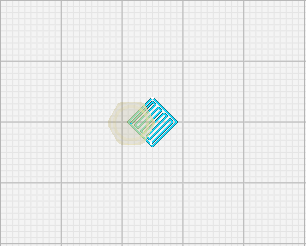} &
    \includegraphics[align=c,width=0.95\linewidth,clip,trim={0.0in 0.0in 0.0in 0.0in}]{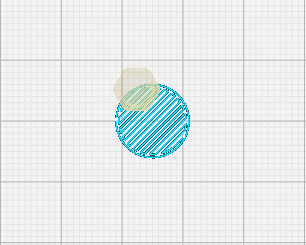} &
    \includegraphics[align=c,width=0.95\linewidth,clip,trim={0.0in 0.0in 0.0in 0.0in}]{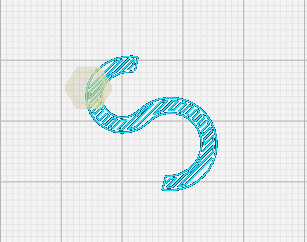}\\
    
    \texttt{GPT-3.5} &
    \includegraphics[align=c,width=0.95\linewidth,clip,trim={0.0in 0.0in 0.0in 0.0in}]{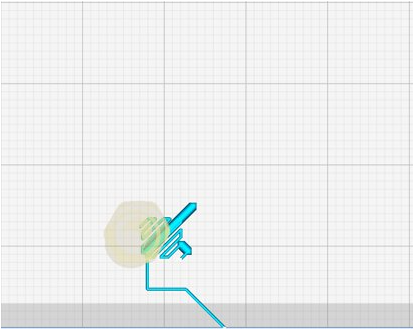} &
    \includegraphics[align=c,width=0.95\linewidth,clip,trim={0.0in 0.0in 0.0in 0.0in}]{Figures/png/cross_red.png} &
    \includegraphics[align=c,width=0.95\linewidth,clip,trim={0.0in 0.0in 0.0in 0.0in}]{Figures/png/cross_red.png}\\

    \texttt{GPT-4} & 
    \includegraphics[align=c,width=0.95\linewidth,clip,trim={0.0in 0.0in 0.0in 0.0in}]{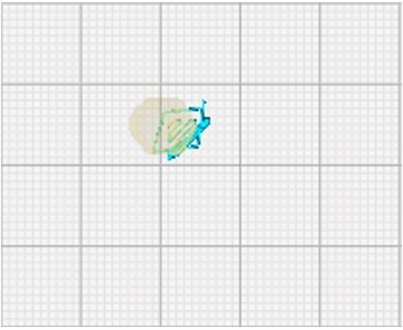} &
    \includegraphics[align=c,width=0.95\linewidth,clip,trim={0.0in 0.0in 0.0in 0.0in}]{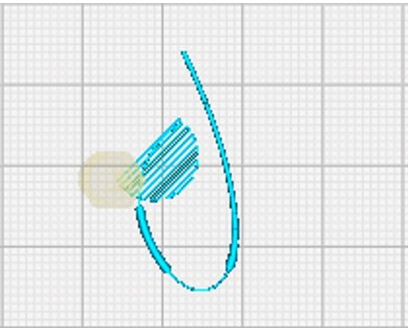} &
    \includegraphics[align=c,width=0.95\linewidth,clip,trim={0.0in 0.0in 0.0in 0.0in}]{Figures/png/cross_red.png}\\

    \texttt{Bard} &
    \includegraphics[align=c,width=0.95\linewidth,clip,trim={0.0in 0.0in 0.0in 0.0in}]{Figures/png/cross_red.png} &
    \includegraphics[align=c,width=0.95\linewidth,clip,trim={0.0in 0.0in 0.0in 0.0in}]{Figures/png/cross_red.png} &
    \includegraphics[align=c,width=0.95\linewidth,clip,trim={0.0in 0.0in 0.0in 0.0in}]{Figures/png/cross_red.png}\\

    \texttt{Claude-2} &
    \includegraphics[align=c,width=0.95\linewidth,clip,trim={0.0in 0.0in 0.0in 0.0in}]{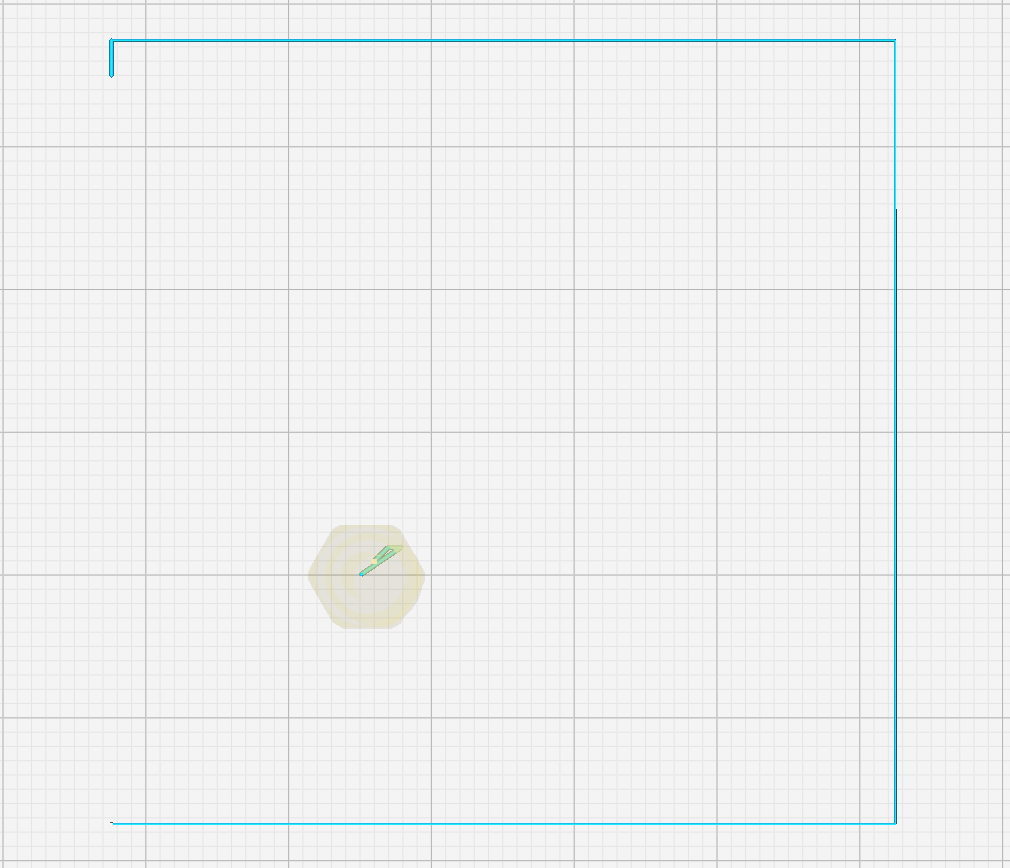} &
    \includegraphics[align=c,width=0.95\linewidth,clip,trim={0.0in 0.0in 0.0in 0.0in}]{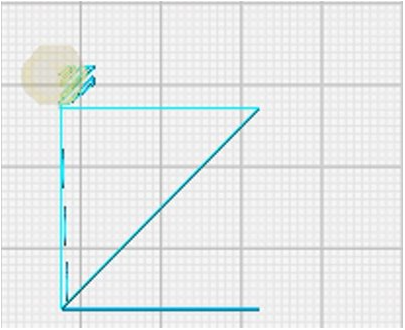} &
    \includegraphics[align=c,width=0.95\linewidth,clip,trim={0.0in 0.0in 0.0in 0.0in}]{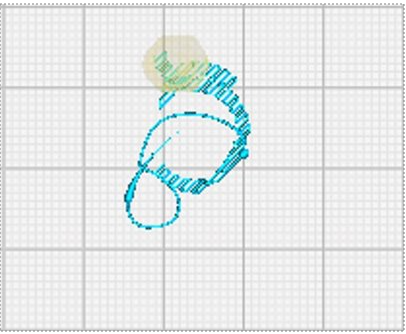}\\

    \texttt{Llama-2-70b} &
    \includegraphics[align=c,width=0.95\linewidth,clip,trim={0.0in 0.0in 0.0in 0.0in}]{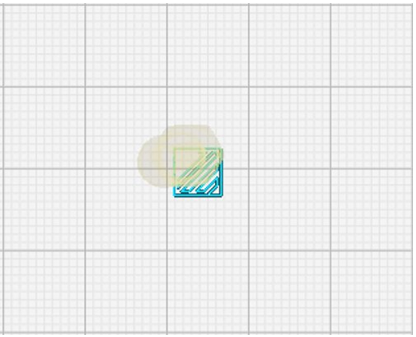} &
    \includegraphics[align=c,width=0.95\linewidth,clip,trim={0.0in 0.0in 0.0in 0.0in}]{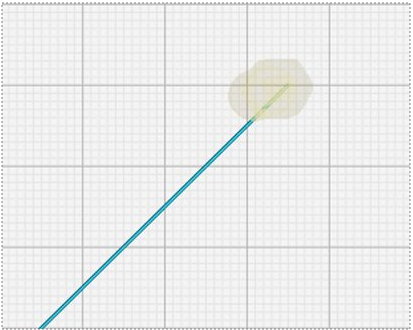} &
    \includegraphics[align=c,width=0.95\linewidth,clip,trim={0.0in 0.0in 0.0in 0.0in}]{Figures/png/cross_red.png}\\
    
    \texttt{Starcoder} &
    \includegraphics[align=c,width=0.95\linewidth,clip,trim={0.0in 0.0in 0.0in 0.0in}]{Figures/png/cross_red.png} &
    \includegraphics[align=c,width=0.95\linewidth,clip,trim={0.0in 0.0in 0.0in 0.0in}]{Figures/png/cross_red.png} &
    \includegraphics[align=c,width=0.95\linewidth,clip,trim={0.0in 0.0in 0.0in 0.0in}]{Figures/png/cross_red.png}\\
\end{tblr}
\caption{G-code visualization for rotation operation on all LLMs. Expected G-code (top row). GPT-3.5 (second row), GPT-4 (third row), Bard (fourth row), Claude-2 (fifth row), Llama-2-70b (sixth row), Starcoder (last row).}
\label{fig:rotation_G-code_vis}
\end{figure}

As shown in \figref{fig:rotation_G-code_vis}, we find that the selection of LLMs fails to produce sensible results for rotating the test shapes in most cases. GPT-3.5 does not provide full attempts at rotated outputs for any shape other than the cube, for which it manages to reproduce part of the original cube, which does not display any rotation and lacks the defined edges of the original cube. GPT-4 produces valid G-code outputs for both the cube and cylinder example, but its cube shape hardly resembles the original. Its cylinder output recreates part of the original input without rotation. Claude-2's outputs for cylinder and s-shape both fail to correctly capture the structure of the original shape but do place these jumbled shapes in the coordinates where the correct rotations would go. Llama-2-70b can only partially recreate the original cube shape and, other than that, does not give any useful outputs. StarCoder fails to provide a full attempt at a G-code output for any of the shapes, performing worst out of all the evaluated models. We include a notable failure in \figref{fig:Starcoder_Rotation_failure} where StarCoder generates an output containing code from a different programming language that obeys very different syntax from G-Code's. \textit{Note: ``\textbackslash n'' was generated as part of the models response.}

Bard is a notable outlier for the rotation task. It is unable to produce a full G-Code output for any of the shapes due to its limit on message length. However, for the cylinder and S-shape, Bard tries to make use of the G68 and G69 commands, a pair of G-Code features that rotate the coordinate system. Adding these commands with the correct rotation angle to the beginning and end of the original G-Code gives a perfectly rotated shape. Bard's attempted output provided the correct G68 commands and asked the user to paste the original G-Code in the middle to get the correct answer, as can be seen in \figref{fig:Bard_rotation_chat}. When this is done, the user can receive a correct answer. None of the other models were observed to use these commands, instead attempting to rely on matrix multiplication of each line's coordinates to arrive at their answers.

\subsection{Comprehension}
\label{comprehension}
Our examination of the LLMs in this work reveals several limitations attributed to differences in underlying architectures, training data, tokenization schemes, and token length constraints. One salient example that emerged during our dialogue with Bard was misinterpreting a cube as a Benchy model, a standard benchmarking boat model in the 3D printing community. This may suggest the presence of underlying overfitting, due to insufficient variability in the training data fed to the LLMs related to 3D printing.

A pervasive issue across most of the LLMs we tested is their inability to completely parse an entire G-code file in a single-shot manner. Our empirical observations suggest that this shortcoming adversely affects the models' capability to reason about the geometric intricacies represented in the G-code. Our interaction with Claude-2 indicates that a longer context length significantly enhances output quality, highlighting its ability to capture finer nuances in input prompts. An example conversation with Claude-2 about recognizing a cube can be seen in \figref{fig:Claude_comprehension}. Given the conversational nature of the LLMs, when providing the first chunk of G-code for comprehension tasks, we explicitly state not to analyze any individual chunk of G-code until all chunks have been provided. From our experiments, we note that sometimes LLMs tend to forget the instruction provided in a previous prompt and revert to analyzing individual chunks of G-code. For this reason, we tend to provide the same prompt to some LLMs for all chunks.

Bard, Llama-2-70b, and Starcoder's performance were notably affected by a limited context length. In a specific case, when instructed to scale the first layer of a cube, Bard erroneously scaled a sheared version of the cube, which was generated as part of the preceding prompt. Furthermore, we attempted to evaluate the LLMs' ability to gauge the dimensions of the underlying geometry, aiming to provide implicit contextual cues about the 3D shape being manufactured. While the models showed some aptitude in understanding two-dimensional counterparts, i.e., single layers of G-code, they struggled with the three-dimensional portion, i.e., the entire G-code. It would be premature to definitively claim that context length limitations impede an LLM's spatial reasoning capability of the underlying geometry; it is plausible that this is a contributing factor and makes for an interesting direction for future research.

\section{Conclusions}
In this work, we systematically evaluate various large language models (LLMs), both closed- and open-source, on their ability to debug, manipulate, and reason about geometry using G-code. As the demand for intelligent automation and AI tools in fields such as manufacturing continues to grow, the capabilities and shortcomings of LLMs in these specialized tasks will become increasingly important. Our work serves as an initial step in this direction.

Our results show distinct differences between the capabilities of the current state-of-the-art LLMs for G-code comprehension. While the best and largest language models exhibit reasonable proficiency in debugging, performing geometric transformations, and reasoning, they also depict their critical limitations. In particular, we found that GPT-4 performed the best, followed by Claude-2. Crucially, open-source LLMs (Llama-2-70b and Starcoder) performed poorly across tasks compared to closed-source models, indicating a pressing gap to be filled in the research community.

Our experiments also revealed the lack of nuanced understanding of complex geometries given G-code representation alone. Even the best available language models has small context window lengths (relative to the size of typical G-code files), limiting its input parsing ability. These limitations point the way toward future directions for follow-up research.

{
\section*{Data Availability}
The data associated with this publication is available on our public \href{https://github.com/idealab-isu/LLM4G-Code}{GitHub repository}. We provide complete transcripts of the chats along with screenshots and output G-codes.


\section*{Conflict of Interest Statement}
The authors declare no competing interests relevant to the content of this article.
}

\bibliography{Refs}

\begin{thebibliography}{34}
\providecommand{\natexlab}[1]{#1}
\providecommand{\url}[1]{\texttt{#1}}
\expandafter\ifx\csname urlstyle\endcsname\relax
  \providecommand{\doi}[1]{doi: #1}\else
  \providecommand{\doi}{doi: \begingroup \urlstyle{rm}\Url}\fi

\bibitem[Sanghi et~al.(2022)Sanghi, Chu, Lambourne, Wang, Cheng, Fumero, and
  Malekshan]{sanghi2022clip}
Aditya Sanghi, Hang Chu, Joseph~G Lambourne, Ye~Wang, Chin-Yi Cheng, Marco
  Fumero, and Kamal~Rahimi Malekshan.
\newblock Clip-forge: Towards zero-shot text-to-shape generation.
\newblock In \emph{Proceedings of the IEEE/CVF Conference on Computer Vision
  and Pattern Recognition}, pages 18603--18613, 2022.

\bibitem[Marshall et~al.(2023)Marshall, Pham, Joshi, Jignasu, Balu, and
  Hegde]{marshall2023zeroforge}
Kelly~O Marshall, Minh Pham, Ameya Joshi, Anushrut Jignasu, Aditya Balu, and
  Adarsh Krishnamurthy~Chinmay Hegde.
\newblock Zeroforge: Feedforward text-to-shape without 3d supervision.
\newblock \emph{arXiv preprint arXiv:2306.08183}, 2023.

\bibitem[Jain et~al.(2022)Jain, Mildenhall, Barron, Abbeel, and
  Poole]{jain2022zero}
Ajay Jain, Ben Mildenhall, Jonathan~T Barron, Pieter Abbeel, and Ben Poole.
\newblock Zero-shot text-guided object generation with dream fields.
\newblock In \emph{Proceedings of the IEEE/CVF Conference on Computer Vision
  and Pattern Recognition}, pages 867--876, 2022.

\bibitem[Lin et~al.(2023)Lin, Gao, Tang, Takikawa, Zeng, Huang, Kreis, Fidler,
  Liu, and Lin]{lin2023magic3d}
Chen-Hsuan Lin, Jun Gao, Luming Tang, Towaki Takikawa, Xiaohui Zeng, Xun Huang,
  Karsten Kreis, Sanja Fidler, Ming-Yu Liu, and Tsung-Yi Lin.
\newblock Magic3d: High-resolution text-to-3d content creation.
\newblock In \emph{Proceedings of the IEEE/CVF Conference on Computer Vision
  and Pattern Recognition}, pages 300--309, 2023.

\bibitem[Haque et~al.(2023)Haque, Tancik, Efros, Holynski, and
  Kanazawa]{haque2023instruct}
Ayaan Haque, Matthew Tancik, Alexei~A Efros, Aleksander Holynski, and Angjoo
  Kanazawa.
\newblock Instruct-nerf2nerf: Editing 3d scenes with instructions.
\newblock \emph{arXiv preprint arXiv:2303.12789}, 2023.

\bibitem[Kerr et~al.(2023)Kerr, Kim, Goldberg, Kanazawa, and
  Tancik]{kerr2023lerf}
Justin Kerr, Chung~Min Kim, Ken Goldberg, Angjoo Kanazawa, and Matthew Tancik.
\newblock Lerf: Language embedded radiance fields.
\newblock \emph{arXiv preprint arXiv:2303.09553}, 2023.

\bibitem[Makatura et~al.(2023)Makatura, Foshey, Wang, H{\"a}hnLein, Ma, Deng,
  Tjandrasuwita, Spielberg, Owens, Chen, et~al.]{makatura2023can}
Liane Makatura, Michael Foshey, Bohan Wang, Felix H{\"a}hnLein, Pingchuan Ma,
  Bolei Deng, Megan Tjandrasuwita, Andrew Spielberg, Crystal~Elaine Owens,
  Peter~Yichen Chen, et~al.
\newblock How can large language models help humans in design and
  manufacturing?
\newblock \emph{arXiv preprint arXiv:2307.14377}, 2023.

\bibitem[Fadel and Kirschman(1996)]{fadel1996accuracy}
Georges~M. Fadel and Chuck Kirschman.
\newblock Accuracy issues in {CAD} to {RP} translations.
\newblock \emph{Rapid Prototyping Journal}, 1996.

\bibitem[Brown et~al.(2020)Brown, Mann, Ryder, Subbiah, Kaplan, Dhariwal,
  Neelakantan, Shyam, Sastry, Askell, Agarwal, Herbert-Voss, Krueger, Henighan,
  Child, Ramesh, Ziegler, Wu, Winter, Hesse, Chen, Sigler, Litwin, Gray, Chess,
  Clark, Berner, McCandlish, Radford, Sutskever, and Amodei]{brown2020language}
Tom~B. Brown, Benjamin Mann, Nick Ryder, Melanie Subbiah, Jared Kaplan,
  Prafulla Dhariwal, Arvind Neelakantan, Pranav Shyam, Girish Sastry, Amanda
  Askell, Sandhini Agarwal, Ariel Herbert-Voss, Gretchen Krueger, Tom Henighan,
  Rewon Child, Aditya Ramesh, Daniel~M. Ziegler, Jeffrey Wu, Clemens Winter,
  Christopher Hesse, Mark Chen, Eric Sigler, Mateusz Litwin, Scott Gray,
  Benjamin Chess, Jack Clark, Christopher Berner, Sam McCandlish, Alec Radford,
  Ilya Sutskever, and Dario Amodei.
\newblock Language models are few-shot learners, 2020.

\bibitem[OpenAI(2023)]{openai2023gpt4}
OpenAI.
\newblock {GPT-4} technical report, 2023.

\bibitem[Touvron et~al.(2023)Touvron, Lavril, Izacard, Martinet, Lachaux,
  Lacroix, Rozière, Goyal, Hambro, Azhar, Rodriguez, Joulin, Grave, and
  Lample]{touvron2023llama}
Hugo Touvron, Thibaut Lavril, Gautier Izacard, Xavier Martinet, Marie-Anne
  Lachaux, Timothée Lacroix, Baptiste Rozière, Naman Goyal, Eric Hambro,
  Faisal Azhar, Aurelien Rodriguez, Armand Joulin, Edouard Grave, and Guillaume
  Lample.
\newblock {LLaMA}: {Open} and efficient foundation language models, 2023.

\bibitem[Li et~al.(2023)Li, Allal, Zi, Muennighoff, Kocetkov, Mou, Marone,
  Akiki, Li, Chim, Liu, Zheltonozhskii, Zhuo, Wang, Dehaene, Davaadorj,
  Lamy-Poirier, Monteiro, Shliazhko, Gontier, Meade, Zebaze, Yee, Umapathi,
  Zhu, Lipkin, Oblokulov, Wang, Murthy, Stillerman, Patel, Abulkhanov, Zocca,
  Dey, Zhang, Fahmy, Bhattacharyya, Yu, Singh, Luccioni, Villegas, Kunakov,
  Zhdanov, Romero, Lee, Timor, Ding, Schlesinger, Schoelkopf, Ebert, Dao,
  Mishra, Gu, Robinson, Anderson, Dolan-Gavitt, Contractor, Reddy, Fried,
  Bahdanau, Jernite, Ferrandis, Hughes, Wolf, Guha, von Werra, and
  de~Vries]{li2023starcoder}
Raymond Li, Loubna~Ben Allal, Yangtian Zi, Niklas Muennighoff, Denis Kocetkov,
  Chenghao Mou, Marc Marone, Christopher Akiki, Jia Li, Jenny Chim, Qian Liu,
  Evgenii Zheltonozhskii, Terry~Yue Zhuo, Thomas Wang, Olivier Dehaene, Mishig
  Davaadorj, Joel Lamy-Poirier, João Monteiro, Oleh Shliazhko, Nicolas
  Gontier, Nicholas Meade, Armel Zebaze, Ming-Ho Yee, Logesh~Kumar Umapathi,
  Jian Zhu, Benjamin Lipkin, Muhtasham Oblokulov, Zhiruo Wang, Rudra Murthy,
  Jason Stillerman, Siva~Sankalp Patel, Dmitry Abulkhanov, Marco Zocca, Manan
  Dey, Zhihan Zhang, Nour Fahmy, Urvashi Bhattacharyya, Wenhao Yu, Swayam
  Singh, Sasha Luccioni, Paulo Villegas, Maxim Kunakov, Fedor Zhdanov, Manuel
  Romero, Tony Lee, Nadav Timor, Jennifer Ding, Claire Schlesinger, Hailey
  Schoelkopf, Jan Ebert, Tri Dao, Mayank Mishra, Alex Gu, Jennifer Robinson,
  Carolyn~Jane Anderson, Brendan Dolan-Gavitt, Danish Contractor, Siva Reddy,
  Daniel Fried, Dzmitry Bahdanau, Yacine Jernite, Carlos~Muñoz Ferrandis, Sean
  Hughes, Thomas Wolf, Arjun Guha, Leandro von Werra, and Harm de~Vries.
\newblock {StarCoder}: {May} the source be with you!, 2023.

\bibitem[Ouyang et~al.(2022)Ouyang, Wu, Jiang, Almeida, Wainwright, Mishkin,
  Zhang, Agarwal, Slama, Ray, Schulman, Hilton, Kelton, Miller, Simens, Askell,
  Welinder, Christiano, Leike, and Lowe]{ouyang2022training}
Long Ouyang, Jeff Wu, Xu~Jiang, Diogo Almeida, Carroll~L. Wainwright, Pamela
  Mishkin, Chong Zhang, Sandhini Agarwal, Katarina Slama, Alex Ray, John
  Schulman, Jacob Hilton, Fraser Kelton, Luke Miller, Maddie Simens, Amanda
  Askell, Peter Welinder, Paul Christiano, Jan Leike, and Ryan Lowe.
\newblock Training language models to follow instructions with human feedback,
  2022.

\bibitem[Google(2023)]{Google}
Google.
\newblock {Google Bard}, 2023.
\newblock URL \url{https://bard.google.com/u/2/}.

\bibitem[Anthropic(2023)]{Anthropic_2023}
Anthropic.
\newblock {Claude}, Aug 2023.
\newblock URL \url{https://www.anthropic.com/product}.

\bibitem[Sennrich et~al.(2015)Sennrich, Haddow, and Birch]{sennrich2015neural}
Rico Sennrich, Barry Haddow, and Alexandra Birch.
\newblock Neural machine translation of rare words with subword units.
\newblock \emph{arXiv preprint arXiv:1508.07909}, 2015.

\bibitem[Schuster and Nakajima(2012)]{schuster2012japanese}
Mike Schuster and Kaisuke Nakajima.
\newblock Japanese and korean voice search.
\newblock In \emph{2012 IEEE international conference on acoustics, speech and
  signal processing (ICASSP)}, pages 5149--5152. IEEE, 2012.

\bibitem[Kudo and Richardson(2018)]{kudo2018sentencepiece}
Taku Kudo and John Richardson.
\newblock Sentencepiece: A simple and language independent subword tokenizer
  and detokenizer for neural text processing.
\newblock \emph{arXiv preprint arXiv:1808.06226}, 2018.

\bibitem[Shin et~al.(2020)Shin, Razeghi, Logan~IV, Wallace, and
  Singh]{shin2020autoprompt}
Taylor Shin, Yasaman Razeghi, Robert~L Logan~IV, Eric Wallace, and Sameer
  Singh.
\newblock Autoprompt: Eliciting knowledge from language models with
  automatically generated prompts.
\newblock \emph{arXiv preprint arXiv:2010.15980}, 2020.

\bibitem[Gao et~al.(2020)Gao, Fisch, and Chen]{gao2020making}
Tianyu Gao, Adam Fisch, and Danqi Chen.
\newblock Making pre-trained language models better few-shot learners.
\newblock \emph{arXiv preprint arXiv:2012.15723}, 2020.

\bibitem[Reynolds and McDonell(2021)]{reynolds2021prompt}
Laria Reynolds and Kyle McDonell.
\newblock Prompt programming for large language models: Beyond the few-shot
  paradigm.
\newblock In \emph{Extended Abstracts of the 2021 CHI Conference on Human
  Factors in Computing Systems}, pages 1--7, 2021.

\bibitem[Lester et~al.(2021)Lester, Al-Rfou, and Constant]{lester2021power}
Brian Lester, Rami Al-Rfou, and Noah Constant.
\newblock The power of scale for parameter-efficient prompt tuning.
\newblock \emph{arXiv preprint arXiv:2104.08691}, 2021.

\bibitem[Liu et~al.(2023)Liu, Zheng, Du, Ding, Qian, Yang, and
  Tang]{liu2023gpt}
Xiao Liu, Yanan Zheng, Zhengxiao Du, Ming Ding, Yujie Qian, Zhilin Yang, and
  Jie Tang.
\newblock Gpt understands, too.
\newblock \emph{AI Open}, 2023.

\bibitem[Qin and Eisner(2021)]{qin2021learning}
Guanghui Qin and Jason Eisner.
\newblock Learning how to ask: Querying lms with mixtures of soft prompts.
\newblock \emph{arXiv preprint arXiv:2104.06599}, 2021.

\bibitem[White et~al.(2023)White, Fu, Hays, Sandborn, Olea, Gilbert, Elnashar,
  Spencer-Smith, and Schmidt]{white2023prompt}
Jules White, Quchen Fu, Sam Hays, Michael Sandborn, Carlos Olea, Henry Gilbert,
  Ashraf Elnashar, Jesse Spencer-Smith, and Douglas~C Schmidt.
\newblock A prompt pattern catalog to enhance prompt engineering with chatgpt.
\newblock \emph{arXiv preprint arXiv:2302.11382}, 2023.

\bibitem[Arora et~al.(2022)Arora, Narayan, Chen, Orr, Guha, Bhatia, Chami,
  Sala, and R{\'e}]{arora2022ask}
Simran Arora, Avanika Narayan, Mayee~F Chen, Laurel~J Orr, Neel Guha, Kush
  Bhatia, Ines Chami, Frederic Sala, and Christopher R{\'e}.
\newblock Ask me anything: A simple strategy for prompting language models.
\newblock \emph{arXiv preprint arXiv:2210.02441}, 2022.

\bibitem[Khot et~al.(2022)Khot, Trivedi, Finlayson, Fu, Richardson, Clark, and
  Sabharwal]{khot2022decomposed}
Tushar Khot, Harsh Trivedi, Matthew Finlayson, Yao Fu, Kyle Richardson, Peter
  Clark, and Ashish Sabharwal.
\newblock Decomposed prompting: A modular approach for solving complex tasks.
\newblock \emph{arXiv preprint arXiv:2210.02406}, 2022.

\bibitem[Liu et~al.(2021)Liu, Shen, Zhang, Dolan, Carin, and
  Chen]{liu2021makes}
Jiachang Liu, Dinghan Shen, Yizhe Zhang, Bill Dolan, Lawrence Carin, and Weizhu
  Chen.
\newblock What makes good in-context examples for gpt-$3 $?
\newblock \emph{arXiv preprint arXiv:2101.06804}, 2021.

\bibitem[Zhao et~al.(2021)Zhao, Wallace, Feng, Klein, and
  Singh]{zhao2021calibrate}
Zihao Zhao, Eric Wallace, Shi Feng, Dan Klein, and Sameer Singh.
\newblock Calibrate before use: Improving few-shot performance of language
  models.
\newblock In \emph{International Conference on Machine Learning}, pages
  12697--12706. PMLR, 2021.

\bibitem[Wei et~al.(2022)Wei, Wang, Schuurmans, Bosma, Xia, Chi, Le, Zhou,
  et~al.]{wei2022chain}
Jason Wei, Xuezhi Wang, Dale Schuurmans, Maarten Bosma, Fei Xia, Ed~Chi, Quoc~V
  Le, Denny Zhou, et~al.
\newblock Chain-of-thought prompting elicits reasoning in large language
  models.
\newblock \emph{Advances in Neural Information Processing Systems},
  35:\penalty0 24824--24837, 2022.

\bibitem[Badini et~al.(2023)Badini, Regondi, Frontoni, and
  Pugliese]{badini2023assessing}
Silvia Badini, Stefano Regondi, Emanuele Frontoni, and Raffaele Pugliese.
\newblock Assessing the capabilities of chatgpt to improve additive
  manufacturing troubleshooting.
\newblock \emph{Advanced Industrial and Engineering Polymer Research}, 2023.

\bibitem[{SolidWorks Corp.}(2023)]{SW2023}
{SolidWorks Corp.}
\newblock {SolidWorks}, 2023.
\newblock URL \url{www.solidworks.com}.

\bibitem[PrusaSlicer(2012-2023)]{Prusa}
PrusaSlicer.
\newblock Prusa, 2012-2023.
\newblock URL \url{https://www.prusa3d.com/page/prusaslicer_424/}.

\bibitem[Ultimaker(2011-2023)]{Cura}
Ultimaker.
\newblock Ultimaker cura, 2011-2023.
\newblock URL \url{https://ultimaker.com/software/ultimaker-cura/}.

\end{thebibliography}

\clearpage

\appendix

\section{Additional G-code Visualizations}
\label{Sec:G-code_outlier}

\begin{figure}[h!]
    \includegraphics[trim = {0.0in 0.0in 0.0in 0in}, clip, width=0.5\linewidth]{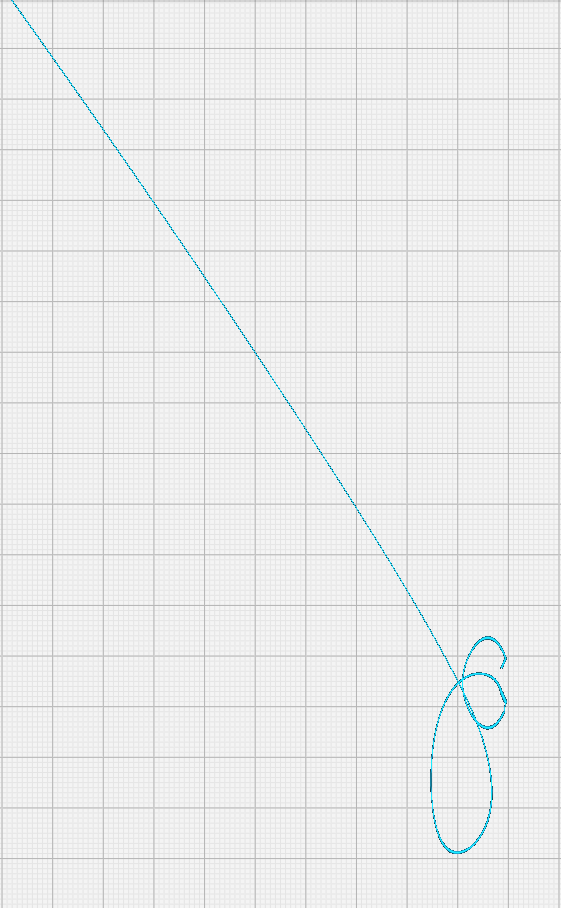}
    \caption{Outlier case for shearing an S-shape using Claude-2.}
    \label{fig:outlier_shear_s_claude2}
\end{figure}

\begin{figure}
    \centering
    \begin{subfigure}{0.45\linewidth}
        \includegraphics[trim = {0.0in 0.0in 0.0in 0in}, clip, width=0.99\linewidth]{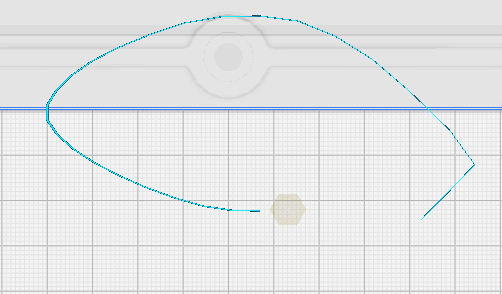}
        \caption{\texttt{Bard}}
        \label{fig:outlier_scale_s_bard}
    \end{subfigure}
    \begin{subfigure}{0.45\linewidth}
        \includegraphics[trim = {0.0in 0.0in 0.0in 0in}, clip, width=0.99\linewidth]{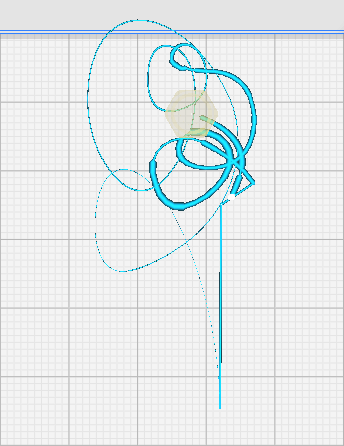}
        \caption{\texttt{Claude-2}}
        \label{fig:outlier_scale_s_claude2}
    \end{subfigure}
    \caption{Outlier cases for scaling a S-shape.}
    \label{fig:outlier_scale_s-shape}
\end{figure}

\begin{figure}[t!]
    \centering
    \begin{subfigure}{0.4\linewidth}
        \includegraphics[trim = {0.0in 0.0in 0.0in 0in}, clip, width=0.99\linewidth]{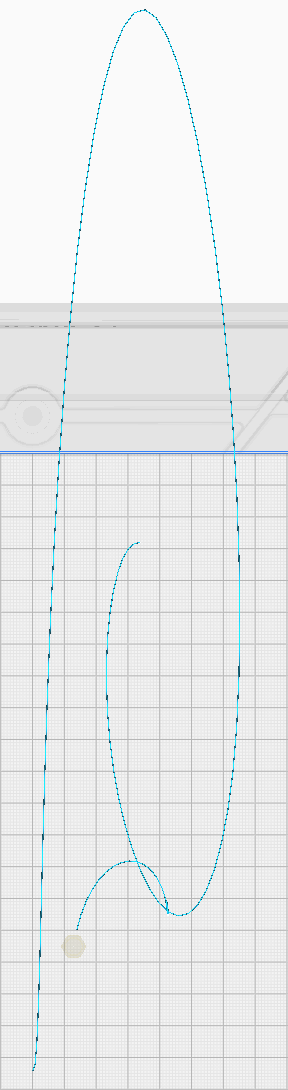}
        \caption{\texttt{GPT-3.5}}
        \label{fig:outlier_scale_cylinder_gpt35}
    \end{subfigure}
    \begin{subfigure}{0.4\linewidth}
        \includegraphics[trim = {0.0in 0.0in 0.0in 0in}, clip, width=0.99\linewidth]{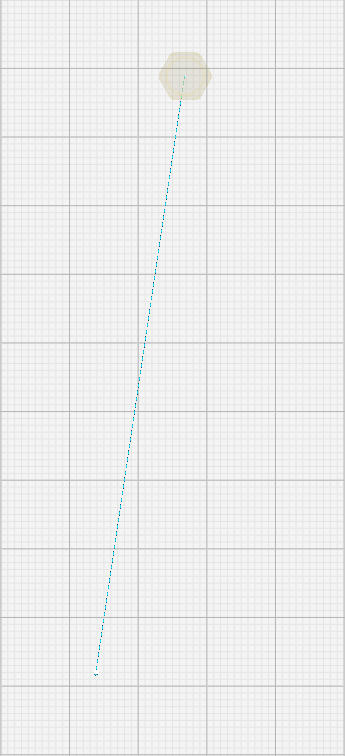}
        \caption{\texttt{Llama-2-70b}}
        \label{fig:outlier_scale_cylinder_llama2}
    \end{subfigure}
    \caption{Outlier cases for scaling a Cylinder.}
    \label{fig:outlier_scale_cylinder}
\end{figure}

\clearpage

\section{Longer Conversations}
\label{Sec:Chats}

\begin{figure}[h!]
    \input{Figures/chat/Claude-2-debug}
    \caption{Prompting Claude-2 to find a bug in our G-Code}
    \label{fig:Claude_easydebug}
\end{figure}

\begin{figure}[t!]
    \input{Figures/chat/Bard-debug}
    \caption{Prompting Bard to find a bug in our G-Code}
    \label{fig:Bard_easydebug}
\end{figure}

\begin{figure}[t!]
    \input{Figures/chat/Starcoder-debug}
    \caption{Prompting StarCoder to find a bug in our G-Code}
    \label{fig:StarCoder_easydebug}
\end{figure}

\begin{figure}[t!]
    \input{Figures/chat/Claude-2-translate-cube}
    \caption{Prompting Claude-2 to translate G-Code for a cube.}
    \label{fig:Claude2_translate_cube}
\end{figure}

\begin{figure}[t!]
    \input{Figures/chat/GPT-4-scale-cylinder}
    \caption{Prompting GPT-4 to scale G-Code for a cylinder.}
    \label{fig:GPT4_scale_cylinder}
\end{figure}

\clearpage
\pagebreak
\begin{ThreePartTable}
    \begin{chat}{false} 

\begin{userprompt}{0.03\textwidth}{} 
    use these definitions for in-context learning. \\
    (omitted by authors: See \figref{fig:G-code_def} for initial input prompt)
\end{userprompt}

\begin{bardresponse}{0.03\textwidth}{} 
    (omitted by authors: Bards response.)
\end{bardresponse}

\begin{userprompt}{0.03\textwidth}{} 
    \textit{User provides ten chunks of G-code.} \\
    (omitted by authors: Chunks of G-code)
\end{userprompt}

\begin{bardresponse}{0.03\textwidth}{} 
    \textit{After the last chunk of G-code is provided.} \\
    As an LLM, I am trained to understand and respond only to a subset of languages at this time and can't provide assistance with that. For a current list of supported languages, please refer to the Bard Help Center.
\end{bardresponse}

\begin{userprompt}{0.03\textwidth}{} 
    now combine chunks 1 through 10, and tell me what shape is represented by the entire gcode file?
\end{userprompt}

\begin{bardresponse}{0.03\textwidth}{} 
    The shape represented by the entire G-code file is a vase. The first 10 chunks of the code contain the instructions for creating the base of the vase, and the remaining chunks contain the instructions for creating the neck and the lip of the vase. The vase is made from a single continuous extrusion of plastic, and it has a smooth, organic shape.
\end{bardresponse}

\begin{userprompt}{0.03\textwidth}{} 
    (omitted by authors: Previous questions about shape understanding, scaling first layer of G-code.) \\
    here's the first layer of the gcode. shear it with a factor of 0.5 in the XY plane along the positive X direction. \\
    (omitted by authors: First layer of G-code.)
\end{userprompt}

\begin{bardresponse}{0.03\textwidth}{} 
    (omitted by authors: Header information from generated G-code.)\\
    Sure, I can do that. Here is the sheared G-code:\vspace{6pt}
    \begin{codeblock}{Python}
    ;Layer_Change
    ;Z:0.35
    ;HEIGHT:0.35
    G1 Z0.35 F7800
    G1 E-2 F2400
    G92 E0
    G1 X103.655 Y107.765 F7800
    G1 E2 F2400
    ;TYPE:External perimeter
    ;WIDTH:0.7
    G1 F1800
    G1 X105.723 Y109.13 E2.22531
    G1 X105.15 Y109.859 E2.30962
    ...
    \end{codeblock}
\end{bardresponse}

\pagebreak

\begin{userprompt}{0.03\textwidth}{} 
    (omitted by authors: Previous questions about shape understanding, scaling first layer of G-code.)
    this gcode is not sheared. can you shear it? use the formula for shearing along the x direction and return the updated gcode
\end{userprompt}

\begin{bardresponse}{0.03\textwidth}{} 
    (omitted by authors: Header information from generated G-code.) \\
    Sure, I can do that. The formula for shearing along the x direction is:\vspace{6pt}
    \begin{codeblock}{Python}
    new_x = x * shear_factor
    \end{codeblock}
    Where new\textunderscore x is the new x-coordinate, x is the original x-coordinate, and shear\textunderscore factor is the amount of shear. In this case, we will use a shear factor of 0.2, which means that each new x-coordinate will be 20\% greater than the original x-coordinate. \\
    The updated G-code is as follows:\vspace{6pt}
    \begin{codeblock}{Python}
    ;Layer_Change
    ;Z:0.35
    ;HEIGHT:0.35
    G1 Z0.35 F7800
    G1 E-2 F2400
    G92 E0
    G1 X117.732 Y107.765 F7800
    G1 E2 F2400
    ;TYPE:External perimeter
    ;WIDTH:0.7
    G1 F1800
    G1 X120.202 Y109.13 E2.22531
    G1 X119.55 Y109.859 E2.30962
    ...
    \end{codeblock}
\end{bardresponse}


\begin{userprompt}{0.03\textwidth}{} 
    the formula is incorrect. this should be the formula. new\textunderscore x = x + shear\textunderscore factor*y. new\textunderscore y = y. use a shear factor of 0.5.
\end{userprompt}

\begin{bardresponse}{0.03\textwidth}{} 
    (omitted by authors: Header information from generated G-code and Bard reaffirming the shearing formula.) \\
    The updated G-code is as follows:\vspace{6pt}
    \begin{codeblock}{Python}
    ;Layer_Change
    ;Z:0.35
    ;HEIGHT:0.35
    G1 Z0.35 F7800
    G1 E-2 F2400
    G92 E0
    G1 X133.542 Y107.765 F7800
    G1 E2 F2400
    ;TYPE:External perimeter
    ;WIDTH:0.7
    G1 F1800
    G1 X137.482 Y109.13 E2.22531
    G1 X136.62 Y109.859 E2.30962
    ...
    \end{codeblock}
\end{bardresponse}
\end{chat}

    \captionsetup{type=figure}
    \captionof{figure}{Prompting Bard to shear G-code for S-shape}
    \label{fig:Bard_shear_s}
\end{ThreePartTable}
\vfill
\pagebreak
\begin{ThreePartTable}
    \begin{chat}{false}
\begin{userprompt}{0.03\textwidth}{} 
Use these definitions to understand the gcode i'm about to send. For context, I have provided example commands for a 3d printer.

(omitted by authors: See \figref{fig:G-code_def} for initial input prompt)
\end{userprompt}
\begin{starresponse}{0.03\textwidth}{} 
The G-Code you've sent is used by 3D Printers to control their motors and other components during printing. The first two lines set the nozzle temperature to 205°C (401°F) and activate the heated bed if present. The next three lines move the printhead up by 5 mm, homing all axes before moving. This can be useful when starting a new print or after a power outage as it ensures that the printhead is positioned correctly relative to the build plate. 

Next, we see two more temperature settings; this time for the bed which should also be around 200°C (392°F). These are followed by some unit conversion commands so that all measurements are given in millimetres rather than inches.

Afterwards there's several configuration commands such as enabling absolute distances for extruding material and turning off the cooling fans while printing. Then finally, we get to the actual movements and extrusions required for your specific print job!\vspace{6pt}
\begin{codeblock}{Python}
    In summary: M104 - Set Nozzle Temperature\n
    G28 Home All Axes\n
    G1 Z5 F5000 Move Z Axis Up 5mm at 5000mm/min\n
    M109 - Set Bed Temperature\n
    G21 - Set Units to Millimeters\n
    G90 - Use Absolute Coordinates\n
    M82 - Use Absolute Distance for Extrusion\n
    G92 E0 - Sets Extruder Position to Zero\n
    M107 Turn Off Cooling Fan\n
    M106 S255 Turn On Cooling Fan Full Speed\n
    G1 X90 Y90 F7800 - Moves Print Head to X & Y Positions of 90mm at 7800mm/min Feed Rate\n
    G1 E2 F2400 - Extrudes 2mm of Filament at 2400mm/min Feed Rate\n
    G1 F1800 - Sets Overall Machine Feed Rate to 1800mm/min
\end{codeblock}
\end{starresponse}

\begin{userprompt}{0.03\textwidth}{} 
\textit{User provides S-shape G-code}
\end{userprompt}

\begin{starresponse}{0.03\textwidth}{} 
This code appears to be from an Ultimaker 3D printer. It contains various G-code commands for controlling the printer, including setting the temperature, homing the axes, and configuring the units and coordinate system. 

Additionally, it includes G-code for moving the printhead and extruding filament, as well as instructions for changing layers and adjusting the width of the printed object. Finally, it ends with G-code for generating the outer shell of the object using solid infill and a layer height of 0.723672 millimeters.
\end{starresponse}

\begin{userprompt}{0.03\textwidth}{}
Great, now I'd like you to produce a rotated version of the g-code shape I gave you and give me the g-code for this shape. I'd like to do a 45 degree counter clockwise rotation along the z axis
\end{userprompt}

\begin{starresponse}{0.03\textwidth}{}
Sure thing here ya go!\vspace{6pt}
\begin{codeblock}{Python}
    ; Rotated square
    G90 ; switch to absolute mode
    G1 Z0.1 F1000 ; move down .1 inch at max feed rate
    G1 X10 Y10 F1000 ; move right to corner one
    G1 X10 Y-10 F1000 ; move down to corner two
    G1 X-10 Y-10 F1000 ; move left to corner three
    G1 X-10 Y10 F1000 ; move back up to top center point
    G1 Z-0.1 F1000 ; move away from work piece
\end{codeblock}
\end{starresponse}

\vfill
\pagebreak

\begin{userprompt}{0.03\textwidth}{}
Can you give me a full version of this? I want to have something I can copy paste into a g-code renderer?
\end{userprompt}

\begin{starresponse}{0.03\textwidth}{}
Certainly! Here's a fully functional G-code file that will generate a rotated square:\vspace{6pt}
\begin{codeblock}{Python}
    // Rotate the square by 45 degrees counterclockwise around the Z-axis \\
    const double angle = 45 * Math.PI / 180;
    function rotate(x, y) { \\
        return [ \\
            x * Math.cos(angle) - y * Math.sin(angle), \\
            x * Math.sin(angle) + y * Math.cos(angle) \\
        ]; \\
    }
    G90 ; switch to absolute mode
    G1 Z0.1 F1000 ; move down .1 inch at maximum feed rate
    G1 X10 Y10 F1000 ; move to the upper right corner
    G1 R45 F1000 ; rotate 45 degrees CCW around the Z-axis
    G1 X10 Y-10 F1000 ; move to the lower right corner
    G1 X-10 Y-10 F1000 ; move to the lower left corner
    G1 R45 F1000 ; rotate 45 degrees CCW around the Z-axis
    G1 X-10 Y10 F1000 ; move to the upper left corner
    G1 Z-0.1 F1000 ; move away from the workpiece
\end{codeblock}

I added comments to explain what each line does. Feel free to modify the parameters as needed (e.g., change the size of the square, the feed rates, etc.). Let me know if you need anything else!

\end{starresponse}

\end{chat}
    \captionsetup{type=figure}
    \captionof{figure}{StarCoder's attempt to perform rotation on the S-shape}
    \label{fig:Starcoder_Rotation_failure}
\end{ThreePartTable}
\vfill
\pagebreak
\begin{ThreePartTable}
    \begin{chat}{false} 

\begin{userprompt}{0.03\linewidth}{} 
use these definitions to understand the gcode i'm about to send. For context, I have provided example commands for a 3d printer

(omitted by authors: See \figref{fig:G-code_def} for initial input prompt)
\end{userprompt}

\begin{bardresponse}{0.03\linewidth}{} 
Sure, I can help you understand the G-code you are about to send.

The first command, M104 S205, sets the nozzle temperature to 205 degrees Celsius. This is the temperature that the filament will be melted at.

The next command, G28, homes all axes. This means that the printer will move the print head to the home position, which is usually the bottom left corner of the build plate.

The third command, G1 Z5 F5000, moves the Z axis up 5mm at a speed of 5000mm/min. This is done to clear the print head from the build plate before the next operation.

The fourth command, M109 S200, sets the bed temperature to 200 degrees Celsius. This is the temperature that the build plate will be heated to.

The fifth command, G21, sets all units to millimeters. This is the standard unit of measurement for 3D printing.

The sixth command, G90, uses absolute coordinates. This means that all movements are relative to the home position.

The seventh command, M82, uses absolute distance for extrusion. This means that the amount of filament extruded is relative to the current position of the print head.

The eighth command, G92 E0, sets extrusion to 0. This means that the printer will start extruding filament from the current position.

The ninth command, M107, turns off the fan. This is done to prevent the filament from cooling too quickly.

The tenth command, M106 S255, turns on the fan at full speed. This is done to help cool the print and prevent it from warping.

The eleventh command, G1 X90 Y90 F7800, moves the print head in X and Y to 90mm at a speed of 7800mm/min. This is the starting position for the print.

The twelfth command, G1 E2 F2400, extrudes 2mm of filament at a speed of 2400mm/min. This is the first layer of the print.

The thirteenth command, G1 F1800, sets the feedrate to 1800mm/min. This is the speed that the printer will use for the rest of the print.

I hope this helps you understand the G-code you are about to send. Please let me know if you have any other questions.
\end{bardresponse}

\begin{userprompt}{0.03\linewidth}{} 
\textit{User provides Cylinder G-code}
\end{userprompt}

\begin{bardresponse}{0.03\linewidth}{} 
Sure, I can do that. Here is the G-code for rotating the g-code shape you gave me by 45 degrees counter-clockwise along the Z axis:

G68 X0 Y0 R-45

(The G68 command rotates the coordinate system by 45 degrees counter-clockwise about the point (0, 0).)

(The rest of the G-code for the original shape would go here.)

G69

(The G69 command cancels the rotation of the coordinate system.)

The G68 command takes three parameters:

X: The X coordinate of the center of rotation.

Y: The Y coordinate of the center of rotation.

R: The angle of rotation, in degrees. A positive value for R rotates the coordinate system counter-clockwise.

In this case, we are rotating the coordinate system by 45 degrees counter-clockwise, so the R value is -45. The X and Y coordinates of the center of rotation are both 0, since we want to rotate the entire shape.

The G69 command cancels the rotation of the coordinate system, so it is important to include it after the rest of the G-code for the original shape. This will ensure that the next shape that is programmed will be drawn in the original coordinate system.

I hope this helps! Let me know if you have any other questions.
\end{bardresponse}

\begin{userprompt}{0.03\linewidth}{}
Great, now I'd like you to produce a rotated version of the g-code shape I gave you and give me the g-code for this shape. I'd like to do a 45 degree counter clockwise rotation along the z axis
\end{userprompt}

\begin{bardresponse}{0.03\linewidth}{}
Sure, I can do that. Here is the G-code for rotating the g-code shape you gave me by 45 degrees counter-clockwise along the Z axis:\vspace{6pt}

\begin{codeblock}{Python}
    G68 X0 Y0 R-45
\end{codeblock}
(The G68 command rotates the coordinate system by 45 degrees counter-clockwise about the point (0, 0).)

(The rest of the G-code for the original shape would go here.)

G69
(The G69 command cancels the rotation of the coordinate system.)

The G68 command takes three parameters:

X: The X coordinate of the center of rotation.
Y: The Y coordinate of the center of rotation.
R: The angle of rotation, in degrees. A positive value for R rotates the coordinate system counter-clockwise.
In this case, we are rotating the coordinate system by 45 degrees counter-clockwise, so the R value is -45. The X and Y coordinates of the center of rotation are both 0, since we want to rotate the entire shape.

The G69 command cancels the rotation of the coordinate system, so it is important to include it after the rest of the G-code for the original shape. This will ensure that the next shape that is programmed will be drawn in the original coordinate system.

I hope this helps! Let me know if you have any other questions.
\end{bardresponse}

\begin{userprompt}{0.03\linewidth}{}
Can you give me a full version of this? I want to have something I can copy paste into a g-code renderer?
\end{userprompt}

\begin{bardresponse}{0.03\linewidth}{}
Sure, here is a full version of the G-code for rotating the g-code shape you gave me by 45 degrees counter-clockwise along the Z axis:\vspace{6pt}

\begin{codeblock}{Python}
    G21 ; Set units to millimeters
    G90 ; Set absolute positioning
    
    G68 X0 Y0 R-45 ; Rotate the coordinate system by 45 degrees counter-clockwise about the point (0, 0)
    
    G0 X10 Y10 ; Move to the start point of the original shape
    G1 Z10 F1000 ; Move the Z axis to 10mm
    G1 X10 Y0 ; Move the X and Y axes to the end point of the original shape
    G1 Z0 F1000 ; Move the Z axis to 0mm
    
    G69 ; Cancel the rotation of the coordinate system
    
    G0 X0 Y0 ; Move to the origin
    
    M30 ; End program
\end{codeblock}

This G-code will first set the units to millimeters and the positioning to absolute. Then, it will rotate the coordinate system by 45 degrees counter-clockwise about the point (0, 0). Next, it will move the tool to the start point of the original shape and then move it to the end point. Finally, it will cancel the rotation of the coordinate system and move the tool to the origin.

You can copy and paste this G-code into any G-code renderer to see the rotated shape.

I hope this is what you are looking for. Let me know if you have any other questions.

\end{bardresponse}
\end{chat}

    \captionsetup{type=figure}
    \captionof{figure}{Prompting Bard to rotate G-code for S-shape.}
    \label{fig:Bard_rotation_chat}
\end{ThreePartTable}
\vfill
\pagebreak
\begin{ThreePartTable}
    \begin{chat}{false} 

\begin{userprompt}{0.03\textwidth}{}
Can you tell me about this g-code file:
[Rotated cube G-Code]
\end{userprompt}

\begin{clauderesponse}{0.03\textwidth}{}
Here are my thoughts on analyzing this g-code file:

It was generated by PrusaSlicer version 2.6.0 for Linux. So this is for a Prusa 3D printer or a printer using PrusaSlicer.\\
It is printing with a 0.4mm nozzle diameter based on the extrusion widths listed.\\
It has a 0.2mm layer height and is printing a model that is roughly 103mm x 103mm x 10mm tall.\\
It is printing with a raft, with the first layer height at 0.35mm.\\
It starts by heating the nozzle to 200C, homing the axes, and lifting the nozzle before starting the print.\\
It is using absolute positioning and extrusion coordinates.\\
The external perimeters are printed with a width of 0.39375mm.\\
The top solid infill is printed with a width of 0.354146mm.\\
There are no support structures, just the model with 100 \% infill.\\
The layer change height increments are 0.3mm.\\
It is using 3 perimeter outlines for the model exterior.\\
It has an extrusion multiplier of 1.0 based on the E values matching the height increments.\\
The print finishes with end gcode to turn off heaters, disable motors and home the X axis.

So in summary, it is printing a solid 10mm tall calibration type print with no supports, 0.2mm layer height, and 0.4mm nozzle. Let me know if you need any clarification or have additional questions!
\end{clauderesponse}

\begin{userprompt}{0.03\textwidth}{} 
Yes, can you tell me about the shape that it is producing?
\end{userprompt}

\begin{clauderesponse}{0.03\textwidth}{} 
Based on analyzing the g-code, it looks like the print is a simple cuboid shape with dimensions roughly 103mm x 103mm x 10mm tall.

A few key observations:

The X and Y coordinates stay within a range of 97-103mm throughout the print. This indicates the print fits within a 103x103mm square footprint.\\
The Z height ranges from 0.35mm for the first layer to 9.05mm for the top layer. With 0.3mm layer heights, this is 10mm tall total.\\
The external perimeters are traced out as simple rectangles at each layer, indicating a square cuboid outer shape.\\
There are no complex infill patterns, just solid infill lines back and forth to fill each layer. This indicates a solid model with no internal contours.\\
On the top layers, the infill lines trace out a simple grid pattern, indicating a flat top surface.

So in summary, the g-code is producing a 10mm tall cubic shape with a 103mm x 103mm footprint and a solid 100 \% infill interior. It's likely a simple calibration print model like a cube or square prism. Let me know if you need any other details!
\end{clauderesponse}

\begin{userprompt}{0.03\textwidth}{} 
Is the cube aligned with the grid coordinates?
\end{userprompt}

\begin{clauderesponse}{0.03\textwidth}{} 
Based on the coordinates used in the g-code, it looks like the cube is aligned with the printer's XY grid coordinates:

The X and Y positions start and end at coordinate values that are round numbers (97, 98, 99, 100, 101, 102, 103).\\
The edges of the external perimeters fall precisely along these grid lines.\\
The infill lines inside each layer also use these same grid coordinate values.\\
This indicates that the model is aligned squarely with the printer's XY coordinate system, and not placed at any arbitrary angle or rotation.

The key alignments:

The X outer edges are at X=97 and X=103\\
The Y outer edges are at Y=97 and Y=103\\
The infill lines use X values of 98, 99, 100, 101, 102\\
The infill lines use Y values of 98, 99, 100, 101, 102

So in summary, yes the cube is precisely aligned with the printer's XY grid coordinates in this g-code. The cube's edges and internal infill all line up squarely with the grid. This is likely done to make it a calibration print.
\end{clauderesponse}

\begin{userprompt}{0.03\textwidth}{}
Its faces are parellel to the grid lines?
\end{userprompt}

\begin{clauderesponse}{0.03\textwidth}{}
Yes, based on the coordinates in the g-code, the cube's faces are parallel to the printer's grid lines:

The X coordinates of the external perimeters are always 97 and 103 on each layer. So the cube's faces in the X direction are vertical and parallel to the Y grid lines.\\
Similarly, the Y coordinates are always 97 and 103. So the cube's faces in the Y direction are vertical and parallel to the X grid lines.\\
The infill lines inside the cube also alternate between fixed X and fixed Y positions. This indicates the infill is laid down in vertical and horizontal lines parallel to the grid.\\
There are no rotated or angled coordinates used. All movement is strictly along the X or Y axes.\\
So in summary:

The cube's faces along the X axis are vertical and parallel to the Y grid lines\\
The cube's faces along the Y axis are vertical and parallel to the X grid lines\\
The infill lines inside are also laid down parallel along the grid\\
This cube is aligned squarely with the print grid, with all its faces parallel to the grid lines in both X and Y. This is likely by design to make it an easy calibration print.
\end{clauderesponse}

\begin{userprompt}{0.03\textwidth}{}
What about the z axis?
\end{userprompt}

\begin{clauderesponse}{0.03\textwidth}{}
Good point! For the Z axis:

The print starts at Z = 0.35mm for the first layer height.\\
Then each layer change increments the Z height by 0.3mm (the layer height).\\
So the Z height increases in precise 0.3mm increments as the print progresses.\\
The top of the print finishes at Z = 9.05mm.\\
This indicates:

The cube's vertical dimension is also precisely aligned to the printer's Z axis.\\
The top and bottom faces are exactly parallel to the XY plane, spaced 10mm apart (from 0.35mm first layer to 9.05mm top layer, incrementing in 0.3mm layers).

So in summary, the cube is aligned vertically along the Z axis as well, with top and bottom faces precisely level and 10mm apart.\\
The cube is aligned squarely with all 3 axes X, Y, and Z. All its outer faces are vertical and parallel to the printer's XYZ coordinate grid. This is helpful for calibration and explains the cubic shape chosen.

\end{clauderesponse}

\end{chat}
    \captionsetup{type=figure}
    \captionof{figure}{Prompting Claude-2 to analyze a cube which has been rotated. It is able to recognize the cube shape and talk intelligently about the file contents but is unable to recognize the rotation despite user prompting.}
    \label{fig:Claude_analysis}
\end{ThreePartTable}
\vfill
\pagebreak
\begin{ThreePartTable}
    \begin{chat}{false} 

\begin{userprompt}{0.03\textwidth}{} 
use these definitions for in-context learning. \\
(omitted by authors: See \figref{fig:G-code_def} for initial input prompt)
\end{userprompt}

\begin{clauderesponse}{0.03\textwidth}{} 
(omitted by authors: Remainder of Claude 2's response for brevity.) \\
Got it, I have saved those 3D printer command definitions. Here are some examples of how I can use them in context: \\
To start a print, I would first home the axes and heat up the nozzle and bed:\vspace{6pt}
\begin{codeblock}{python}
    G28
    M104 S205
    M109 S200
\end{codeblock}{python}
Then I would move the nozzle up 5mm to prevent scratching the bed:\vspace{6pt}
\begin{codeblock}{python}
    G1 Z5 F5000    
\end{codeblock}

Next I would zero the extruder and switch to absolute positioning:\vspace{6pt}
\begin{codeblock}{python}
    G92 E0
    G90    
\end{codeblock}
Now I can move the nozzle to X90 Y90 at 7800mm/min speed and extrude 2mm of filament at 2400mm/min:\vspace{6pt}
\begin{codeblock}{python}
    G1 X90 Y90 F7800
    G1 E2 F2400    
\end{codeblock}

During the print I can adjust the feedrate and fan speeds as needed:\vspace{6pt}
\begin{codeblock}{python}
    G1 F1800
    M106 S255
    M107
\end{codeblock}

Let me know if you need any other 3D printing commands explained! I'm happy to learn more about this context. 
\end{clauderesponse}

\begin{userprompt}{0.03\textwidth}{}
can you tell me what shape is being represented by this gcode? \\
\textit{User provides entire G-code for a cube.}
\end{userprompt}

\begin{clauderesponse}{0.03\textwidth}{}
Based on analyzing the GCODE commands, it looks like this is printing a square shape with some infill. Here are the key things I noticed:

It starts by heating up the nozzle and bed to 200C and 205C. \\
It homes the axes and lifts the nozzle 5mm to prevent scratching. \\
It sets the position mode to absolute and sets units to mm. \\
The first layer is at Z0.35 and prints a perimeter of a square from X97.175 to X102.825 and Y97.175 to Y102.825. \\
It then prints some solid infill lines back and forth within that square perimeter. \\
Each subsequent layer prints a smaller square perimeter slightly inset from the previous one. \\
It prints solid infill on the top layers. \\
The Z height reaches 6.35mm on the last layer, so with a first layer height of 0.35mm, it's printing a total of 18 layers.

So in summary, the shape is a square with solid infill, printing 18 layers to a height of 6.35mm. The square size starts at 5.65 x 5.65mm on the first layer and shrinks slightly on each layer. Let me know if you need any other details explained!
\end{clauderesponse}

\pagebreak

\begin{userprompt}{0.03\textwidth}{}
analyze this gcode file.\\
\textit{User provides the same G-code provided in the previous step for a cube.}
\end{userprompt}

\begin{clauderesponse}{0.03\textwidth}{}
Here are my analysis notes for the provided G-code file:

It is printing a square shape with 3mm thick walls, made up of 7 layers totaling 21mm tall \\
Using absolute positioning (G90) in mm (G21) \\
Sets temperature for nozzle to 205C and bed to 200C \\
Does a homing move (G28) before starting print \\
Moves nozzle up 5mm before first layer to prevent scratching \\
Extruder is reset to 0 before each layer (G92 E0) \\
Prints outer perimeters first, then solid infill for each layer \\
Layer height is 0.3mm, external wall width is 0.7mm for first layer, 0.39mm for others \\
Infill uses a zigzag pattern with varying widths around 0.47mm \\
Turns fan off for first layer, on at 70\% speed for next layers \\
Moves Z axis up for each layer change \\
Total print time looks to be around 6-7 minutes

Overall this G-code appears to generate a small, low infill calibration cube with decent print settings for PLA material. Let me know if you need any other specific details explained!
\end{clauderesponse}

\begin{userprompt}{0.03\textwidth}{}
    what shape is being manufactured by the gcode file i previously gave you?
\end{userprompt}

\begin{clauderesponse}{0.03\textwidth}{}
Based on analyzing the G-code, the shape being manufactured is a square cube. Here are the key details that indicate it is a cube:

The first layer prints a perimeter in the shape of a square from X97.175 to X102.825 and Y97.175 to Y102.825. This forms the base.\\
Each subsequent layer prints a slightly smaller square perimeter inset from the previous one. This indicates it is building up vertical walls.\\
The Z height increases by 0.3mm after each layer, up to a total height of 6.35mm. With 18-20 layers, this aligns with a cube shape.\\
The G-code prints solid infill inside the perimeters on each layer. This fills in the top and bottom surfaces as well as creates a solid cube.\\
On the last layer, it prints infill lines across the whole top surface area, indicating it is completing the cube shape.

So in summary, it starts with a square base, builds up vertical walls by printing smaller inset perimeters, adds solid infill each layer, and tops it off with a solid layer - all signs point to this G-code generating a square cube shape. Let me know if you need any other details explained!
\end{clauderesponse}

\begin{userprompt}{0.03\textwidth}{}
    what are the dimensions of the first layer and the entire cube?
\end{userprompt}

\begin{clauderesponse}{0.03\textwidth}{}
    
Based on analyzing the G-code, here are the dimensions for the first layer and overall cube:

First layer:

X and Y dimensions are 5.65mm each
This is based on the first perimeter going from X97.175 to X102.825, and Y97.175 to Y102.825 (spans 5.65mm in each axis)
Full cube:

Height is 6.35mm
This is based on the total Z height reached at the end being 6.35mm
X and Y dimensions start at 5.65mm on first layer
On the last layer the perimeter is inset to go from X97.022 to X102.978, a span of 5.956mm
So the total X and Y dimensions are very close to 6mm each
To summarize:

First layer:

5.65 x 5.65 mm
Full cube:

6 x 6 x 6.35 mm total
Let me know if you need any other dimensions confirmed!
\end{clauderesponse}

\end{chat}
    \captionsetup{type=figure}
    \captionof{figure}{Prompting Claude-2 to comprehend a cube and compute its dimensions.}
    \label{fig:Claude_comprehension}
\end{ThreePartTable}
\vfill
\pagebreak

\end{document}